\documentclass[useAMS,usenatbib]{mn2e}
\usepackage{natbib}
\usepackage{graphicx}

\input{psfig}
\def\ga{\mathrel{\mathchoice {\vcenter{\offinterlineskip\halign{\hfil
$\displaystyle##$\hfil\cr>\cr\sim\cr}}}
{\vcenter{\offinterlineskip\halign{\hfil$\textstyle##$\hfil\cr>\cr\sim\cr}}}
{\vcenter{\offinterlineskip\halign{\hfil$\scriptstyle##$\hfil\cr>\cr\sim\cr}}}
{\vcenter{\offinterlineskip\halign{\hfil$\scriptscriptstyle##$\hfil
\cr>\cr\sim\cr}}}}}
\def\la{\mathrel{\mathchoice {\vcenter{\offinterlineskip\halign{\hfil
$\displaystyle##$\hfil\cr<\cr\sim\cr}}}
{\vcenter{\offinterlineskip\halign{\hfil$\textstyle##$\hfil\cr<\cr\sim\cr}}}
{\vcenter{\offinterlineskip\halign{\hfil$\scriptstyle##$\hfil\cr<\cr\sim\cr}}}
{\vcenter{\offinterlineskip\halign{\hfil$\scriptscriptstyle##$\hfil
\cr<\cr\sim\cr}}}}}

\title{Vlasov-Poisson in 1D for initially cold systems: post-collapse Lagrangian perturbation theory}

\author[S. Colombi]{St\'ephane
  Colombi\thanks{E-mail: colombi@iap.fr} \\
\\
Institut d'Astrophysique de Paris, CNRS UMR 7095 and UPMC, 98bis, bd Arago, F-75014 Paris, France}
\begin{document}
\voffset -1cm
\date{\today}
\pagerange{\pageref{firstpage}--\pageref{lastpage}} \pubyear{2014}
\maketitle
\label{firstpage}
\begin{abstract}
We study analytically the collapse of an initially smooth, cold, self-gravitating collisionless system in one dimension. The system is described as a central ``${\cal S}$'' shape in phase-space surrounded by a nearly stationary halo acting locally like a harmonic background on the ${\cal S}$. To resolve the dynamics of the ${\cal S}$ under its self-gravity and under the influence of the halo, we introduce a novel approach using post-collapse Lagrangian perturbation theory. This approach allows us to follow the evolution of the system between successive crossing times and to describe in an iterative way the interplay between the central ${\cal S}$ and the halo.  Our theoretical predictions are checked against measurements in entropy conserving numerical simulations based on the waterbag method. While our post-collapse Lagrangian approach does not allow us to compute rigorously the long term behavior of the system, i.e. after many crossing times, it explains the close to power-law behavior of the projected density observed in numerical simulations. Pushing the model at late time suggests that the system could build at some point a very small flat core, but this is very speculative. This analysis shows that understanding the dynamics of initially cold systems requires a {\em fine grained} approach for a correct description of their very central part. The analyses performed here can certainly be extended to spherical symmetry.
\end{abstract}
\begin{keywords}
gravitation -- methods: analytical -- methods: numerical
\end{keywords}

\section{Introduction}
It is currently widely admitted that the matter content of the Universe is dominated by a dark component. Dark matter can be assimilated to a collisionless, self-gravitating fluid, so its dynamics follows Vlasov-Poisson equations. In the one dimensional case that we examine in this work, these equations are given, in the proper units, by 
\begin{equation}
\frac{\partial f}{\partial t}+v \frac{\partial f}{\partial v} - \frac{\partial \phi}{\partial x} \frac{\partial f}{\partial v}=0,
\end{equation}
\begin{equation}
\frac{\partial^2 \phi}{\partial x^2} = 2 \rho(x,t),
\end{equation}
\begin{equation} 
\rho(x,t) \equiv  \int f(x,v',t) {\rm d}v',
\end{equation}
where $x$ is the position, $v$ the velocity, $t$ the time, $f(x,v,t)$ the phase-space density distribution function, $\phi(x,t)$ the gravitational potential and $\rho(x,t)$ the projected density. 

In the current paradigm of large scale structure formation, the favored model supposes that dark matter was initially cold, with close to infinitely small velocity dispersion, that is, in one dimension, 
\begin{equation}
f(x,v,t=0) \equiv \rho(x)\ \delta_{\rm D}[v-v_{\rm ini}(x)].
\label{eq:inif}
\end{equation}
Numerical simulations show that cold dark matter aggregates in dense and compact halos that seem to follow a nearly universal, quasi-steady projected density profile \citep[see, e.g.][and references therein]{NFW1,NFW2,Navarro2010,Diemand2011}. Despite many attempts to understand the origin of this universality, it is not yet clear how this profile, in particular its central structure, builds up. 

It is believed that self-gravitating systems evolve towards steady state after a strong mixing phase designated by {\em violent relaxation} during which the entropy of the system augments \citep[see, e.g.][]{LyndenBell,Tremaine1986}. One common way to try predicting the structure of the system after relaxation consists in using a statistical approach combined with entropy maximization, such as in the theory of \cite{LyndenBell} and its numerous extensions \citep[see, e.g., the recent investigations by][for the case of dark matter halos, but this list is far from exhaustive]{Hjorth2010,Pontzen2013,Carron2013}. However, this approach requires some level of coarse graining of the distribution function, an operation which is not unique nor free of biases \citep[see, e.g.,][]{Chavanis2005,Arad05}. Furthermore, the concept of entropy is not necessarily well defined in the continuous limit \citep[see, e.g.,][and references therein]{Tremaine1986,Chavanis2006}. One important purpose of this article is to convince indirectly the reader that coarse graining cannot be performed everywhere, particularly in zones where the structure of the system presents non trivial singularities that require to be followed dynamically in their full detail, that is in an entropy conserving fashion preserving the very needed memory of initial conditions. 

An alternative approach consists in investigating the subset of self-similar solutions, which allows one to derive important properties of the system without any need, {\em a priori}, for coarse graining \citep[see, e.g.,][]{Fillmore1984,Bertschinger1985,Alard2013}. The major issue with  self-similarity, although being an intuitive outcome of gravitational dynamics, is to be able to find the dynamical route that builds it up. So far, except for important clues provided by perturbation theory \citep[see, e.g.,][]{Moutarde91}, the actual establishment of self-similarity, when it exists, has never been rigorously demonstrated, to my knowledge, otherwise than by numerical experiments.

In this paper, we aim to study the dynamics of a self-gravitating system without using a statistical approach nor assuming self-similarity, but rather by relying on a simplified description of the dynamics capturing all the important physical processes at play.  To do this we focus on the simple case of one dimensional gravity and try to compute analytically the evolution of a single, initially cold and smooth structure, following the footsteps of many previous works in the literature \citep[see, e.g.,][]{Doroshkevich1980,Melott1983,Shandarin1989,Rouet1990,Gurevich1995,Binney,Schulz2013,CT12}.

Our main tool is Lagrangian perturbation theory \cite[see, e.g.][and references therein]{Zeldo,Shandarin1989,Buchert92,Bouchet92,Buchert93,Bouchet95,ptreview} where the small parameter is the displacement field. Lagrangian perturbation theory can be used to follow, at least partly, the evolution of a system beyond collapse time \citep[see, e.g.,][for an example of application]{Pichon99}. However, the strong feedback due to the appearance of singularities at trajectory crossings is usually not taken into account in the calculations except by using some ansatz such as, for instance, the Burgers' equation \citep[see, e.g.,][]{Gurbatov89} or other alternatives \citep[see, e.g.,][and references therein]{Sahni95}. These ans\"atze unfortunately provide a poor description of the internal structure of collapsed structures. The iterative procedure relying on a transport equation of the gravitational field proposed by \cite{Buchert2006} is potentially one exception, but it was not yet tested against numerical simulations. 

The idea of {\em post-collapse Lagrangian perturbation theory} introduced in this work is to follow the system beyond collapse in a ballistic fashion, compute the gravitational field as a function of time in this approximation, then correct the equations of motion accordingly in order to improve the description of the post-crossing dynamics. In this respect, this approach is similar to the iterative procedure advocated by Buchert, except that I propose here to stop at the first iteration.  If one stays sufficiently close to the center of the system, the coordinates of the curve representing it can be approximated by a polynomial at third order in the Lagrangian position, a ``${\cal S}$'' shape \citep[see, e.g.,][]{Shandarin1989}, which makes possible the analytical calculation of the force exerted on any point of the system even when it is multivalued (at most three-valued in our approximation). While our perturbative approach can theoretically be accurate only shortly after shell-crossing, we shall see in fact that it is a spectacularly good approximation up to next crossing time, which will permit us to follow the system during multiple orbits in an iterative fashion.

This paper is thus organized as follow. In \S~\ref{sec:zeldo} we give the initial set up and the solution until first crossing time, which is exactly given by leading order Lagrangian perturbation theory, the so-called Zel'dovich approximation \citep{Zeldo}. Section \ref{sec:postlag} is the core of the paper: it details our post-collapse Lagrangian perturbative approach. It is divided in five parts. The first one describes the way we summarize the system, which is supposed to be composed of a central self-gravitating ${\cal S}$ shape plunged in a close to stationary halo that contributes as a harmonic background on the dynamics of the ${\cal S}$. In the second part, we study the different phases of the motion, in particular whether the local flow is mono- or multi-valued, and compute the gravitational force exerted exerted on an element of fluid of the system as a function of time. Once the acceleration is provided, one can estimate the corresponding corrections on the ballistic motion: this is the subject of third part of \S~\ref{sec:postlag}. Fourth and fifth parts expose the way we propose to iterate the procedure from crossing time to crossing time and discuss the results obtained with it. Section \ref{sec:compnum} compares our analytical predictions to numerical simulations. Finally, Section~\ref{sec:conclusion} summarizes the main results of this article, relates them to maximum entropy methods and comments on possible extensions to higher number of dimensions.

\section{Initial set-up and evolution until first crossing time}
\label{sec:zeldo}
We examine a particular but still quite generic case in equation (\ref{eq:inif}),  with $v(x)=0$ and where the projected density $\rho$ is an even quadratic function of $x$ parametrized as follows,
\begin{equation}
\rho(x) = {\bar \rho}_0 (1-3\alpha x^2),  \quad {\bar \rho}_0 > 0, \quad \alpha > 0, 
\end{equation}
and becomes null when $|x| > x_{\rm M}$ with $x_{\rm M} \leq 1/\sqrt{3 \alpha}$. 

Up to the first crossing time, resolving equations of motion is straightforward in the cold case. To do so, it is convenient, following the footsteps of \cite{Shandarin1989}, to trace the phase-space distribution function with a curve $[x(q,t),v(q,t)]$ where $q$ is a Lagrangian coordinate along the curve. Our initial configuration can be conveniently written as follows:
\begin{eqnarray}
 x(q,t=0) &=& q,\\
 v(q,t=0) &=& 0.
\end{eqnarray}
Again, remind that the system extends in the region
\begin{equation}
q \in [-q_{\rm M},q_{\rm M}], \quad q_{\rm M} \leq 1/\sqrt{3 \alpha}.
\end{equation}

The Lagrangian equations of motion, prior to shell crossing, read
\begin{eqnarray}
\frac{\partial {x}}{\partial t} & = & v, \\
\frac{\partial {v}}{\partial t} & = & -{\rm sgn}(q) M(q),
\end{eqnarray}
with interior mass given by
\begin{equation}
M(q)=2{\bar \rho}_0 |q-\alpha q^3|.
\label{eq:mofq}
\end{equation}

Consider the new variables 
\begin{equation}
{\tilde t}=\sqrt{{\bar \rho}_0}t, \quad {\tilde q}=\sqrt{\alpha} q, \quad
{\tilde x}=\sqrt{\alpha}{ x},\quad {\tilde v}=\frac{\partial {\tilde x}}{{\partial}{\tilde t}}.
\end{equation} 
Then, we simply have
\begin{equation}
\frac{{\partial}{\tilde v}}{{\partial}{\tilde t}}=-{\rm sgn}({\tilde q}) {\tilde M}({\tilde q}),
\label{eq:accvelt}
\end{equation}
with
\begin{equation}
{\tilde M}({\tilde q})=2|{\tilde q}-{\tilde
  q}^3|=\frac{\sqrt{\alpha}}{{\bar \rho}_0} M(q).
\label{eq:accmasst}
\end{equation}
This shows that we can set the parameters 
\begin{equation}
\alpha \equiv {\bar \rho}_0 \equiv 1,
\end{equation}
without any loss of generality, as assumed from now on.

Prior to shell crossing, the solution of the equations of motion
simply follows Zel'dovich dynamics \citep{Zeldo,Shandarin1989} and reads
\begin{eqnarray}
x(q,t) &= &q-(q-q^3)t^2, \label{eq:xdyn}\\
v(q,t) & = & -2(q-q^3)t,\label{eq:vdyn}
\end{eqnarray}
for $|q| \leq q_{\rm M}$.
The projected density can be written
\begin{equation}
\rho(x,t)=\rho(q)\left| \frac{\partial x}{\partial q}
\right|^{-1}=\frac{1-3q^2}{1-t^2+3q^2 t^2}.
\label{eq:projdens}
\end{equation}
The first crossing, defined by,
\begin{equation}
\frac{\partial x(q,t_{\rm c})}{\partial q}|_{q=0} \equiv0,
\end{equation}
happens at {\rm collapse time},
\begin{eqnarray}
t_{\rm c} \equiv t_{{\rm c},1}=1.
\end{eqnarray}
At collapse time, we have
\begin{eqnarray}
x(q,t_{\rm c})=q^3,
\end{eqnarray}
hence, we obtain the well known singular behavior \citep{Zeldo,Arnold1982}
\begin{equation}
\rho(x,t_{\rm c})=\frac{1}{3x^{2/3}}-1.
\label{eq:rhosing}
\end{equation}

\section{From one crossing time to another: post collapse Lagrangian perturbation theory}
\label{sec:postlag}
\begin{figure}
\centerline{\psfig{file=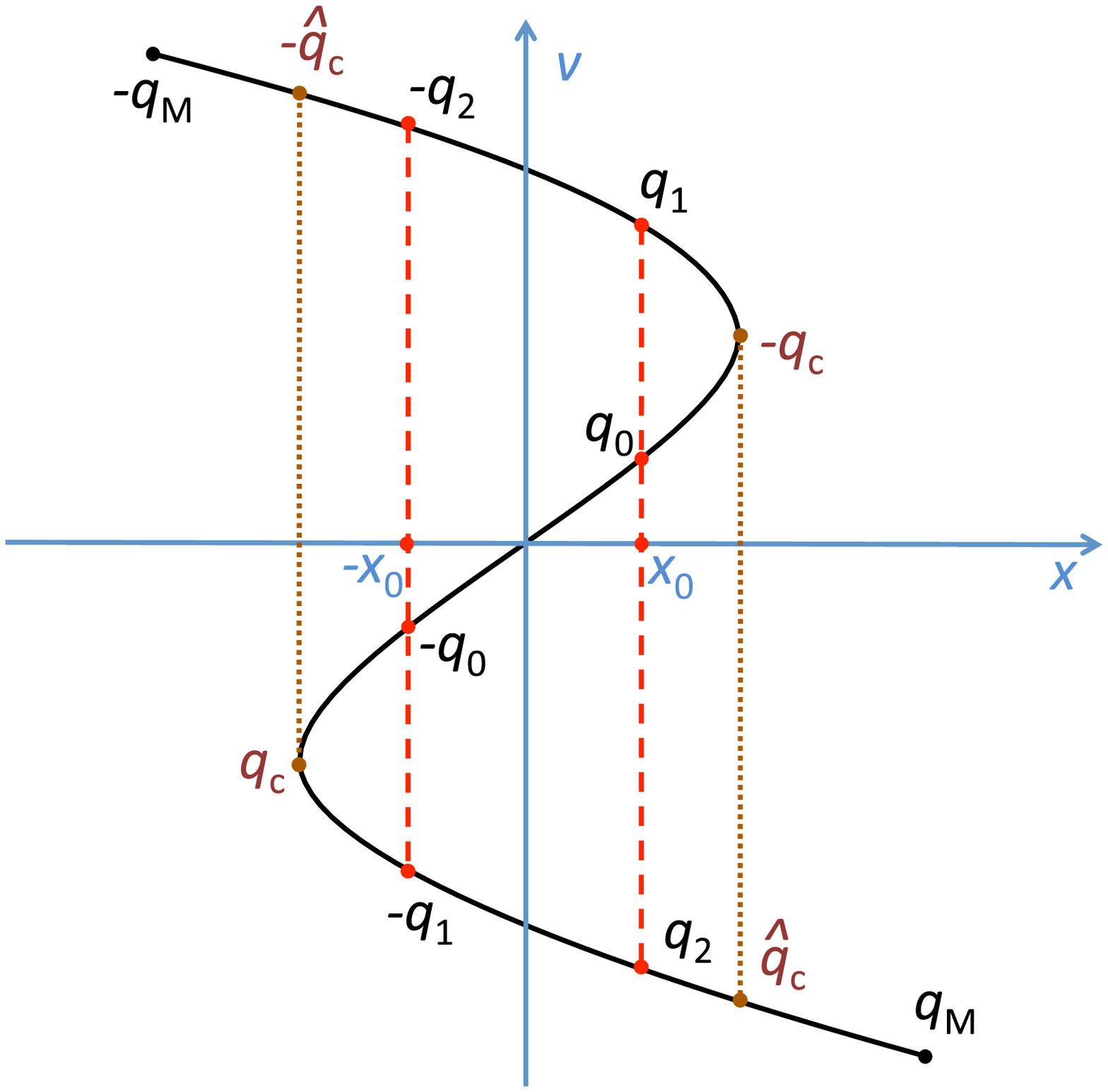,width=8.6cm}}
\caption[]{Schematic representation of the phase-space structure of the system shortly after collapse. The are some points for which the equation
  $x(q,h)=x_0$ has 3 solutions, $(q_0,q_1,q_2)$, with $|q_0| \leq |q_1| \leq |q_2|$. This happens for $|x_0| \leq |x[q_{\rm c}(t),t]|$, with $q_{\rm c}$ defined by equation~(\ref{eq:defqc}). In this case, the mass interior to $x_0$ reads
$M_{\rm int}(|x| \leq x_0)=M(q_2)-M(q_1)+M(q_0)$, where function $M(q)$ is given
by equation~(\ref{eq:mofq}). This is true as long as ${\hat q}_{\rm
    c}(h) \equiv 2 q_{\rm c}(h)\leq q_{\rm M}$.  When ${\hat q}_{\rm c}(h) > q_{\rm M}$, a
small additional complication arises from the fact that Lagrangian
coordinates with $|q| > q_{\rm M}$ do not contribute to the mass.}
\label{fig:explain}
\end{figure}

We notice from equations (\ref{eq:xdyn}) and (\ref{eq:vdyn}) that $x(q,t)$ and $v(q,t)$ are odd third order polynomials of $q$, hence the curve $[x(q,t),v(q,t)]$ is an evolving ``${\cal S}$'' shape. In particular, at collapse time, which corresponds to the first crossing time,
\begin{eqnarray}
x(q,t_{\rm c}) &=& a q^3,  \label{eq:crossingx}\\
v(q,t_{\rm c}) &=& -b q + c q^3, \label{eq:crossingy}
\end{eqnarray}
where  $a$, $b$ and $c$ are positive numbers:
\begin{eqnarray}
a \equiv a_1 = 1, \quad b \equiv b_1 = 2, \quad c \equiv c_1 = 2.
\end{eqnarray}
Beyond collapse time, the evolution of the system becomes more complex than in previous paragraph because there are multiple valued points (Fig.~\ref{fig:explain}), i.e.  the equation  $x(q,t)=x_0$ has either no solution, one solution, 2 or 3 solutions according to the value of
$x_0$ considered. After multiple crossing times, $t \geq t_{{\rm c}, n}$, the system builds up a spiral structure in phase-space, of which the dynamics is difficult to study without resorting to additional assumptions, such as for instance self-similarity \citep{Gurevich1995,Alard2013}. However the central part of this spiral should still remain a ${\cal S}$ as long as the evolution of the system is quiescent. 

In this section, we use a Taylor expansion of the curve $[x(q,t),v(q,t)]$ at third order in $q$ in order to be able to follow the evolution of this ${\cal S}$ shape after each crossing time. Thus, at each crossing time, $t_{\rm c}=t_{{\rm c},n}$, equations (\ref{eq:crossingx}) and (\ref{eq:crossingy}) apply again but with new values of the parameters $a=a_n$, $b=b_n$, $c=c_n$.  This approximation allows one to resolve easily the problem of multiple valued points and this is made possible only because we restrict to third order in $q$. Fortunately, it will be shown in \S~\ref{sec:numexp} that third order is enough to preserve all the important aspects of the dynamics of the center of the system in phase-space. For large enough values of $|q|$ corresponding to the tails of the ${\cal S}$, our approximation will fail, but we shall provide arguments to show that these tails feed a halo around the ${\cal S}$ corresponding, at the fine level, to the spiral mentioned above. In addition, the system is supposed to be immersed in a constant density background, $\rho_{\rm b}=\rho_{{\rm b},n}$, which accounts for the presence of this halo. 

In order to resolve the equations of the dynamics, in particular to compute the force as a function of time, we first write the motion in the presence (or not) of the background $\rho_{\rm b}$,  then compute the additional perturbation induced by the self-gravity of the ${\cal S}$. This perturbation supplies a correction to the evolution of the system valid at second order in time. Even though this means that our post-collapse calculations are only valid during a short time after a crossing, we shall assume that they stand until next crossing, which will allows us to compute $a_{n+1}$, $b_{n+1}$ and $c_{n+1}$ as functions of $a_n$, $b_n$ and $c_{n}$, hence to follow in a {\em fine grained} fashion the dynamics of the center of the system at successive crossing times. 

\subsection{Equations of motion: the perturbative set-up}
In the presence of a pure homogeneous background density $\rho_{\rm b}$, the system follows the harmonic oscillator motion:
\begin{equation}
\frac{\partial^2 x}{\partial t^2}=-\omega x,
\label{eq:x2harmo}
\end{equation}
with 
\begin{equation}
\omega=\sqrt{2 \rho_{\rm b}}.
\end{equation}
With the initial conditions given by equations (\ref{eq:crossingx}) and (\ref{eq:crossingy}), the solution  of equation (\ref{eq:x2harmo}) is
\begin{eqnarray}
x_{\rm b}(q,h) & = & a q^3 \cos( \omega h ) + (-b q + c q^3) \sin( \omega h)/\omega, \label{eq:harmox} \\
v_{\rm b}(q,h) &= & (-b q + c q^3) \cos( \omega h) - a \omega q^3 \sin( \omega h), \label{eq:harmoy}
\end{eqnarray}
with
\begin{equation}
h\equiv t-t_{\rm c}.
\end{equation}
The pure ballistic regime (no background density) is obtained by taking the limit $\omega \rightarrow 0$.
Note importantly that harmonic oscillator dynamics preserves the cubic nature of the curve $[x(q,t),v(q,t)]$ and represents a case where restricting to third order in $q$ is {\em exact}. Basically, the harmonic motion corresponds to a  clockwise rotation of the ${\cal S}$ shape in phase-space. 

Now, we aim to compute the perturbation induced by self-gravity of the system on equations (\ref{eq:harmox}) and (\ref{eq:harmoy}), 
\begin{eqnarray}
x(q,h) &= &x_{\rm b}(q,h)+g(q,h), \label{eq:xharper} \\
v(q,h) & = & v_{\rm b}(q,h)+\frac{\partial g}{\partial h}, \label{eq:vharper}
\end{eqnarray}
where function $g$ is meant to be evaluated at third order in $q$.
To do this, we first calculate the gravitational force exerted by the ${\cal S}$ on a particle of mass unity,
\begin{equation}
F \equiv \frac{\partial^2 g}{\partial h^2},
\end{equation}
assuming that the motion is dominated by the $(x_{\rm b},v_{\rm b})$ contribution, i.e. by setting $g=0$ in equations (\ref{eq:xharper}) and (\ref{eq:vharper}). This allows us, after successive integrations over time on the force, to compute $g$ and its derivative. At early time, one expects $g(q,h) \sim {\cal O}(h^2)$. Our calculation remains accurate as long as the perturbation induced by self-gravity on $(x_{\rm b},v_{\rm b})$ remains small. This should indeed be the case during a full orbit (the trajectory of a matter element between two crossing times) if the background density is large, i.e. $2\ell \rho_{\rm b}=  \ell \omega^2 \gg M$ where $\ell$ is the size of the system and $M$ its mass. In general, though, our approximation will be rigorously valid only shortly after collapse. This is particularly true in the pure ballistic case, $\omega \rightarrow 0$, where the cumulative perturbation induced by $F$ changes completely the trajectory and allows for a new crossing that would never be possible otherwise. In this case, it would be needed to reiterate the procedure on the newly calculated $(x,v)$ to obtain new corrections beyond second order in time. However, such an iteration would not only be extremely cumbersome to perform, it would also be nonsensical, unless considering simultaneously higher order than third in $q$. Nevertheless, we shall see in \S~\ref{sec:numexp} by comparisons with a numerical experiment that the simplified procedure we decided to follow here catches, even quantitatively, all the important features of the dynamics of the center of the system. 

\subsection{The different events during an orbital time and the acceleration as a function of time}
\label{sec:phases}
In this section, we assume, following the discussion of previous paragraph, that the trajectory of a particle tracing the phase-space distribution function is well approximated by the harmonic motion, or, when $\omega \rightarrow 0$, the ballistic solution. We aim to describe, during this trajectory, the important time events that imply different behaviors for the force exerted on the particle. Here, we sketch the most important steps of the calculations. Explicit expressions are given in Appendix~\ref{app:cal}, including series expansions allowing one to simplify the calculations and to perform them at third order in $q$. 

From now on we assume that the particle has initially positive Lagrangian coordinate, $q \geq 0$, but the same reasoning applies symmetrically to $q \leq 0$.  

Before shell-crossing, the flow is monovalued, so the acceleration of the particle induced by the self gravity of the ${\cal S}$ is just given equations (\ref{eq:accvelt}) and (\ref{eq:accmasst}) that we repeat here for clarity with the current notations:
\begin{eqnarray}
F &=&-{\rm sgn}(q) M(q), \label{eq:accvelt2}\\
M(q) &=& 2|q-q^3|. \label{eq:mqsimp}
\end{eqnarray}
During the motion, the ${\cal S}$ shape rotates clockwise in phase-space.  A first event corresponds to the moment $h={\hat h}_{\rm c}(q)$ when the particle meets the handle of the ${\cal S}$ that lies in the positive half-velocity space on Fig.~\ref{fig:explain}. At this time, we have $q={\hat q}_{\rm c}$ on Fig.~\ref{fig:explain} and 
\begin{equation}
x_{\rm b}(q,{\hat h}_{\rm c})=x_{\rm b}[-{q}_{\rm c}({\hat h}_{\rm c}),{\hat h}_{\rm c}], 
\end{equation}
where $q_{\rm c}(h)$ defines the magnitude of the Lagrangian position of the most right part of the handle and therefore solves the following equation: 
\begin{equation}
\frac{{\partial x}_{\rm b}}{\partial q_{\rm c}} \equiv 0.
\label{eq:defqc}
\end{equation}
In our third-order approximation level, 
\begin{equation}
{\hat q}_{\rm c}(h)=2 q_{\rm c}(h).
\label{eq:defhatqc}
\end{equation}
Then ${\hat h}_{\rm c}(q)$ is obtained by resolving the following equation
\begin{equation}
{\hat q}_{\rm c}[{\hat h}_{\rm c}(q)]\equiv q. 
\end{equation}
From this point, the flow becomes three-valued at position $x_0=x(q,h)$. This equation has three solutions:
\begin{equation}
x(q_j,h) \equiv x(q,h), \quad |q_0(q,h)| < |q_1(q,h)| < |q_2(q,h)|,
\end{equation} 
one of them coinciding with $q$, while the two others can be obtained by solving a second order polynomial in $q$, as can be easily seen from equation (\ref{eq:harmox}). At the beginning, we have $q_2=q$, then $q_1=q$ after crossing of the axis $x=0$, corresponding to a sign change of the force exerted on the particle. The phases where $q_2=q$ and $q_1=q$ correspond to a first regime where the acceleration increases with time, until the particle reaches the leftmost part of the handle of the ${\cal S}$ which lies in the negative half-velocity space in Fig.~\ref{fig:explain}, $q=-q_{\rm c}$. This event takes place at time $h=h_{\rm c}(q)$ defined by
\begin{equation}
q_{\rm c}[h_{\rm c}(q)]\equiv q. 
\end{equation}
When $h \geq h_{\rm c}(q)$, we have $q_0=q$, and the force exerted on the particle presents a new regime.

During the phase when the system is multivalued, $h \geq {\hat h}_{\rm c}(q)$, the expression of the acceleration can be easily calculated from the mass of the three portions of the ${\cal S}$ with $|x| \leq |x_0|$:
\begin{eqnarray}
F &=&-{\rm sgn}(q) [M[{\rm min}(|q_2|,q_{\rm M})]-M[{\rm min}(|q_1|,q_{\rm M})] \nonumber \\
 & & \quad \quad \quad \quad +M(q_0)],
\label{eq:massintformula}
\end{eqnarray}
where function $M(q)$ is given by equation (\ref{eq:mqsimp}).  In expression (\ref{eq:massintformula}), we took into account of the finite extent of the system, that is the possibility of $|q_2|$ or both $|q_1|$ and $|q_2|$ to be larger than $q_{\rm M}$. The events $|q_2|=q_{\rm M}$ and $|q_1|=q_{\rm M}$ define two respective moments $h_-(q) \leq h_+(q)$:
\begin{eqnarray}
|q_2(q,h_-)| & \equiv & q_M, \\
|q_1(q,h_+)| & \equiv & q_M.
\end{eqnarray}
When $h_-(q) \leq h \leq h_+(q)$, the tails of the ${\cal S}$ contribute less and less to the the force, until $h = h_+(q)$, which corresponds to the moment when the system becomes monovalued again, that is with an acceleration given again by equation (\ref{eq:accvelt2}).  This obviously complicates the analyses, as it introduces new regimes on the acceleration. While the ordering ${\hat h}_{\rm c}(q) \leq \{ {\hat h}_{\rm c}(q), h_-(q) \} \leq h_+(q)$ is always correct, we have ${\hat h}_{\rm c}(q) \leq h_-(q)$ if $q \leq q_{\rm M}/2$ and the opposite otherwise. 

\begin{figure}
\centerline{\psfig{file=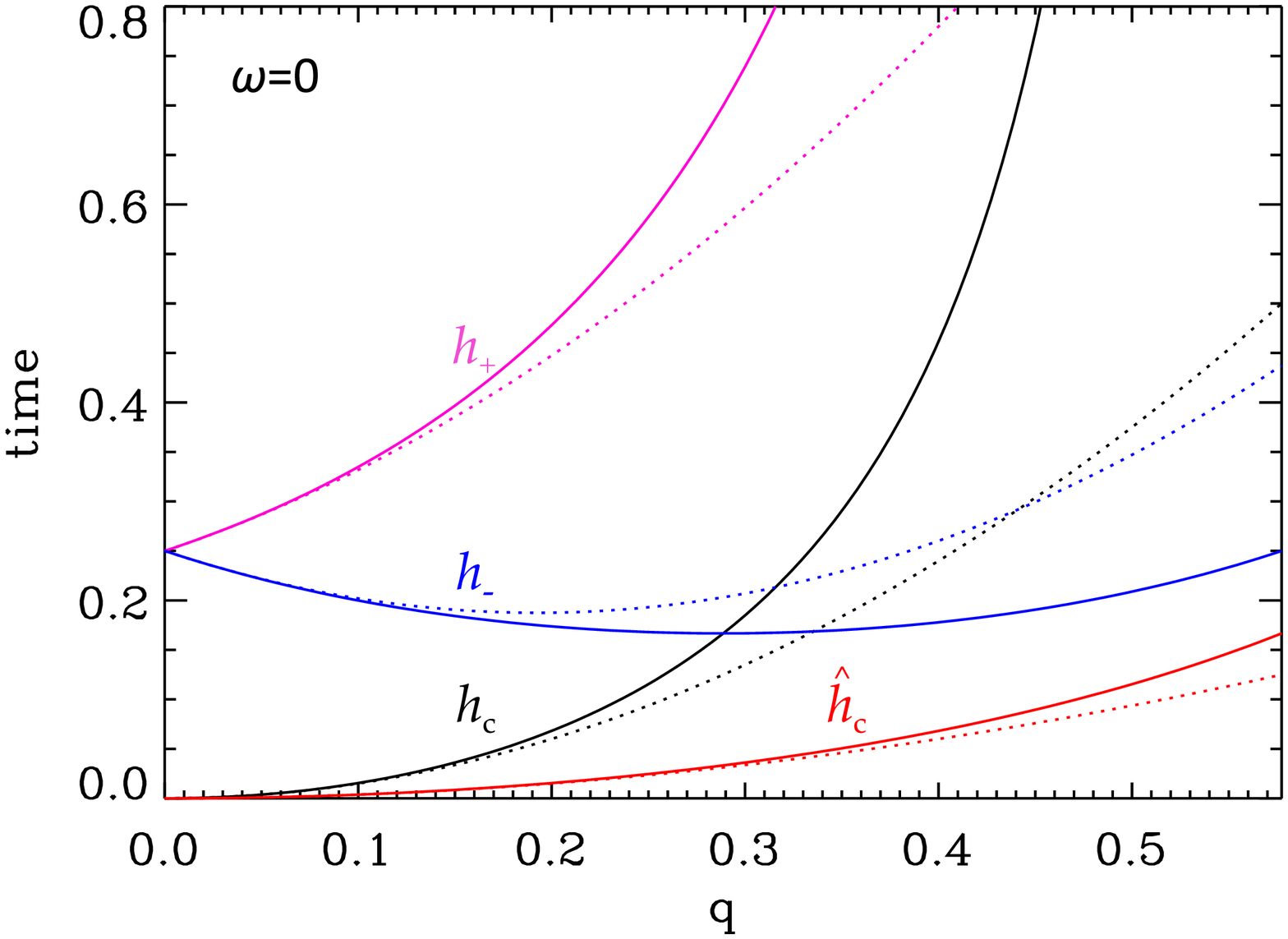,width=8.6cm}}
\centerline{\psfig{file=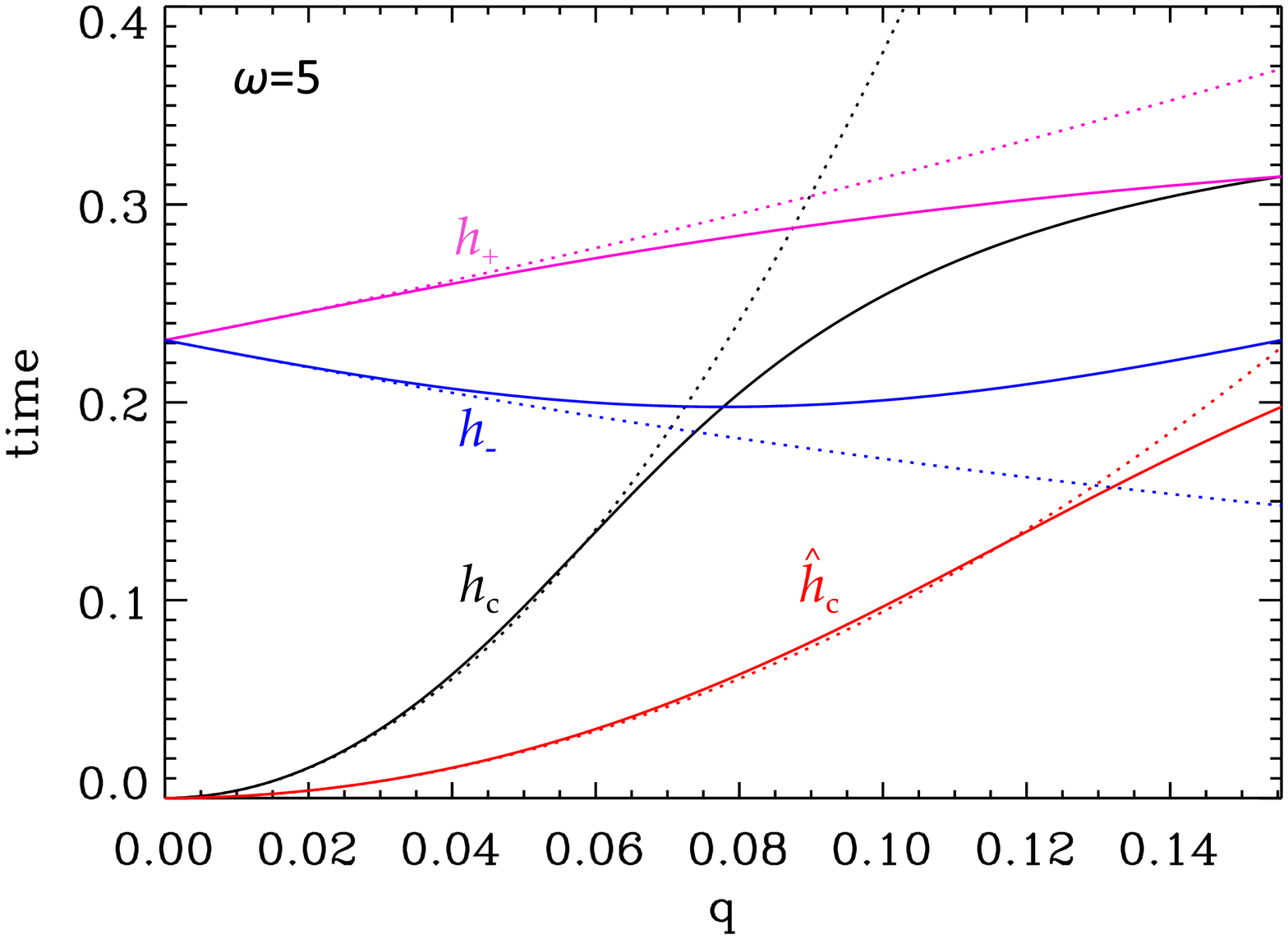,width=8.6cm}}
\caption[]{The critical event times as functions of Lagrangian coordinate $q$. They mark the transitions between different regimes for a particle following the trajectory defined by equations (\ref{eq:harmox}) and (\ref{eq:harmoy}). The solid curves correspond to the exact expressions, equations (\ref{eq:hcexp}), (\ref{eq:hathcexp}) and (\ref{eq:hpm}) of Appendix~\ref{app:timesdetails}, while the dotted ones are the second order expansion in $q$ we use in our calculations. The top panel corresponds to a no density background case, $\omega=0$, with  $a=1$, $b=2$, $c=2$ and $q_{\rm M}=1/\sqrt{3}$, as obtained after first crossing time $t_{{\rm c},1}$, while the bottom one assumes a rather dense background $\omega=5$, with  $ a=17.7$, $b=1.4$, $c=19.4$ and $q_{\rm M}=\sqrt{b/(3c)}$ as discussed later. First crossing time happens at ${\hat h}_{\rm c}(q)$. For ${\hat h}_{\rm c}(q) \leq h= t-t_{{\rm c},1} \leq h_{\rm c}(q)$ the particle is of kind ``$q_2$'' in Fig.~\ref{fig:explain}, becomes of kind ``$q_1$'' when $x_{\rm b}(q,h)$ changes sign and then of kind ''$q_0$'' when $h \geq h_{\rm c}(q)$. The two other times displayed on the figure mark the moment when only one point $q_1 \neq q$ of the ${\cal S}$ shape has the same position  $x_{\rm b}$ as point $q$ ($|q_2| > q_{\rm M}$), that is when $h > h_{-}(q)$, and finally when the system becomes monovalued again ($|q_1|, |q_2| > q_{\rm M}$), when $h > h_{+}(q)$. Note that the intersection of the black and the blue curves, $h_{\rm c}(q)=h_-(q)$, is obtained for $q=q_{\rm M}/2$. }
\label{fig:charactimes}
\end{figure}

\begin{figure}
\centerline{\psfig{file=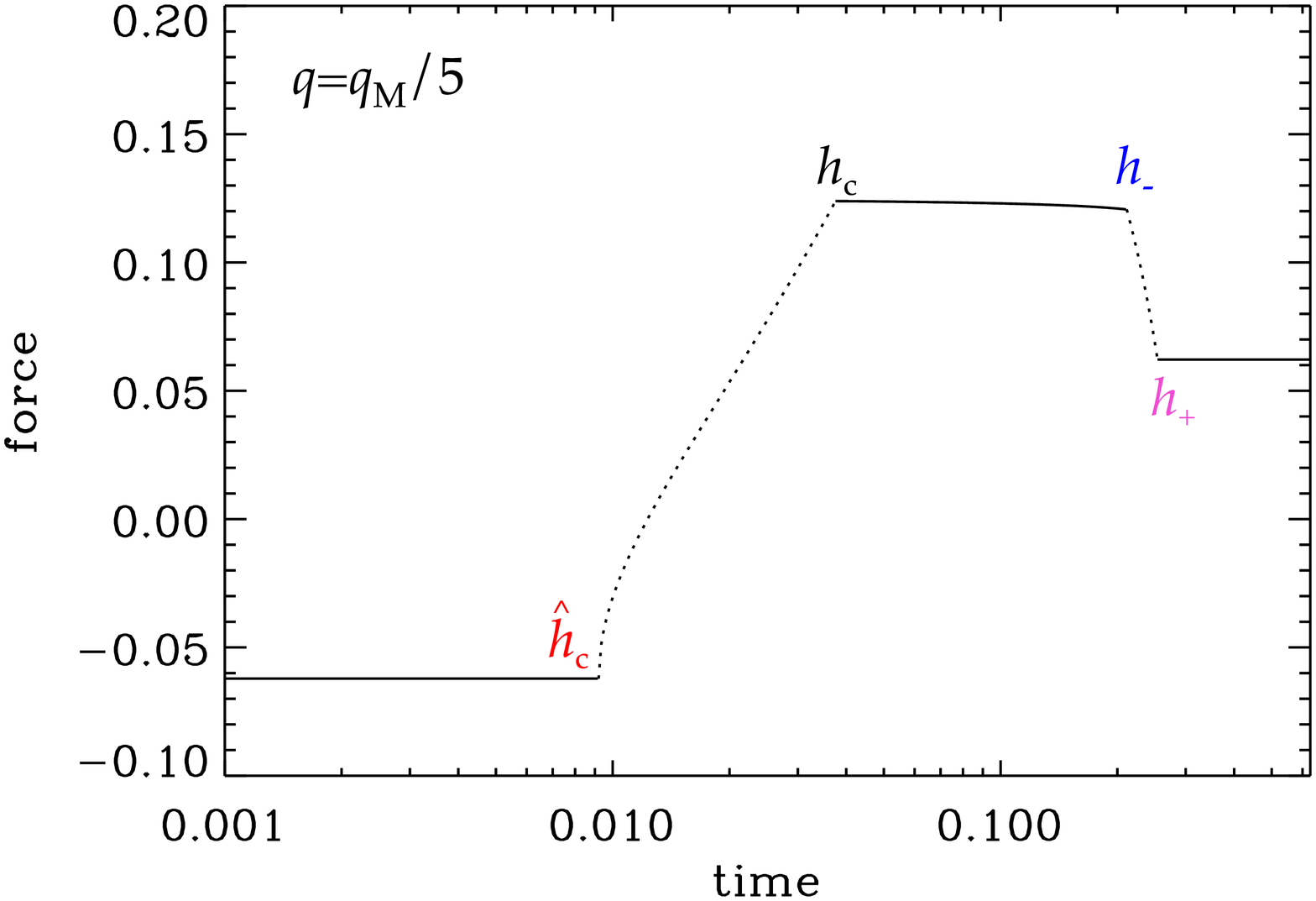,width=8.6cm}}
\centerline{\psfig{file=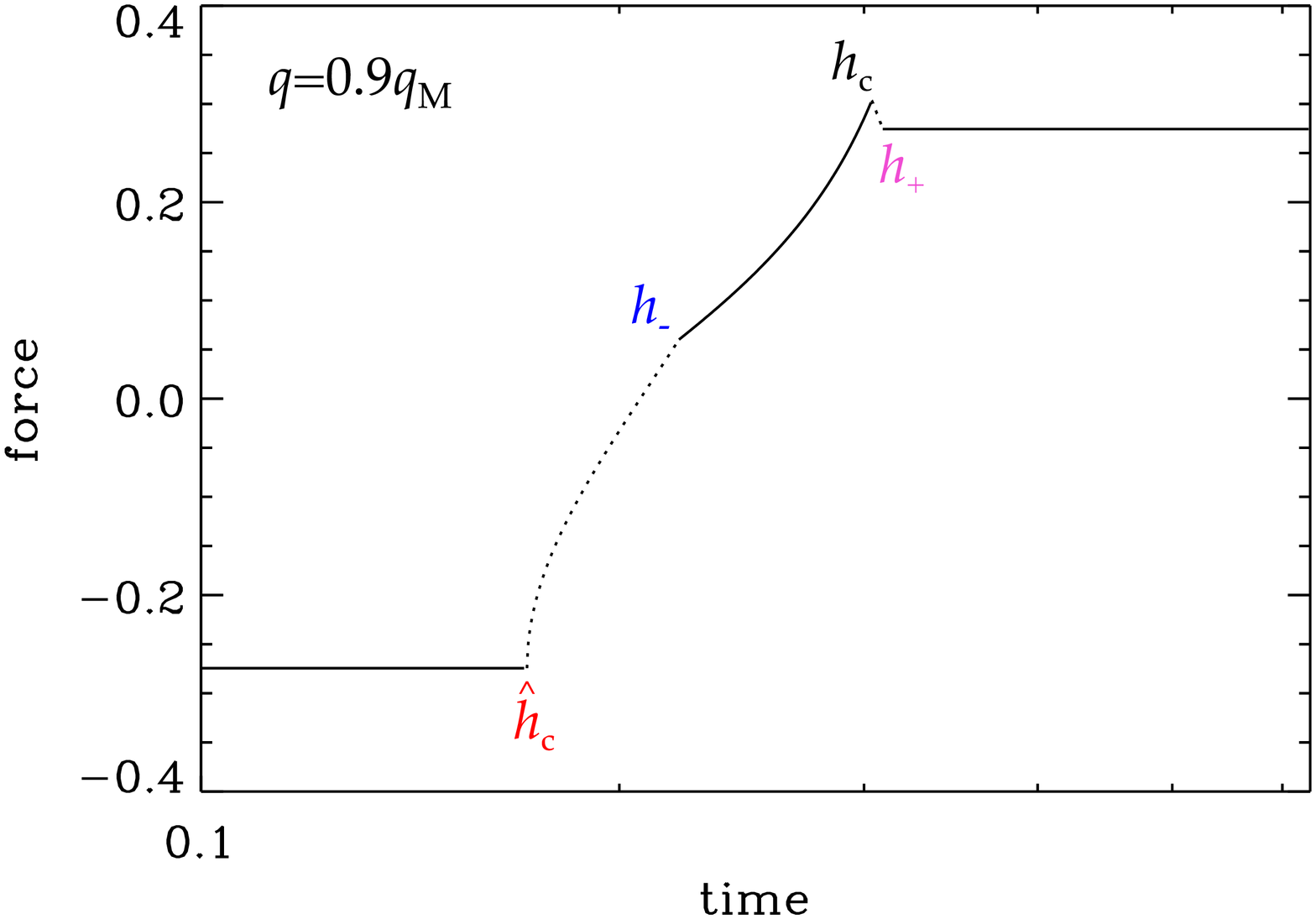,width=8.6cm}}
\caption[]{The self-induced gravitational force exerted by the ${\cal S}$ as a function of time on an element of mass belonging to it, assuming the equations of motion (\ref{eq:harmox}) and (\ref{eq:harmoy}). 
The top panel corresponds to a particle near the center of the ${\cal S}$ while the bottom one correspond to a particle in one of the tails of the ${\cal S}$. The force computed here (see Appendix~\ref{sec:forcecalc} for details)  thus does not include the effect of the harmonic background, which is already accounted for in the equations of motion. The parameters used are the same as in second panel of Fig.~\ref{fig:charactimes}. To distinguish clearly between successive phases, the curves on each panel are alternatively continuous and dotted. The maximum time displayed corresponds to a full harmonic orbit, $h=\pi/\omega$. In the top panel, we have $\int_0^{\rm \pi/\omega} F(q,h) v(q,h) {\rm d}h=-2.7\times10^{-4}$, that implies at net loss of energy during the orbit, while this integral is equal to $1.6\times 10^{-3}$ in the bottom panel, which implies a net gain of energy, as discussed in the main text.}
\label{fig:forcetime}
\end{figure}

To understand better what happens during the trajectory of a matter element, Figure~\ref{fig:charactimes} shows the various event times, ${\hat h}_{\rm c}$, $h_{\rm c}$ and $h_{\pm}$ as functions of $q$, as well as their second order series expansion in $q$ that are needed for subsequent calculations.
This figure is supplemented by Fig.~\ref{fig:forcetime} which displays the force exerted on a point mass particle as a function of time. 

Let us recapitulate again the different phases to which a particle is subject during one orbital time. First we consider the case $|q| \leq q_{\rm M}/2$ (top panel of Fig.~\ref{fig:forcetime}): 
\begin{enumerate}
\item{\em  Prior to first-crossing}, $h \leq {\hat h}_{\rm c}(q)$: the force is simply given by equation (\ref{eq:accvelt2}), exactly as in \S~\ref{sec:zeldo}.
\item{\em Between first-crossing and interior phase}, ${\hat h}(q)
  \leq h \leq h_{\rm c}(q)$: the particle is of kind $q_2$, then $q_1$ when  $x_{\rm b}(q,h)$ changes sign. The force is, from now on, given by equation (\ref{eq:massintformula}) but where $q_{\rm M}$ can be ignored.  Because the particle passes through the center of the system, the force first decreases in magnitude with time, changes sign, then increases in magnitude to reach at the end of this phase its maximum strength, $F[q,h_{\rm c}(q)] =4 q (1-4 q^3)$ (equation \ref{eq:F3} of Appendix~\ref{sec:forcecalc}), roughly {\em twice larger} than prior to shell-crossing. Indeed, not only the part of the ${\cal S}$ with $|q'| \leq q$ contributes to the interior mass, but also the two tails of the ${\cal S}$ ($q_1$ and $q_2$ on Fig.~\ref{fig:explain}).
\item{\em Interior phase}, $h_{\rm c}(q) \leq h \leq h_-(q)$: the particle becomes of kind $q_0$ i.e. ``interior'' to the ${\cal S}$ shape. The force decreases slowly with time as the particle continues orbiting. During this phase, the potential well is deeper than prior to shell-crossing and the particle is losing energy to the tails of the ${\cal S}$. 
\item{\em interior phase, but decreasing contribution of the tails}, $h_-(q) \leq h \leq h_+(q)$: $q_{\rm M}$ cannot be ignored anymore in equation (\ref{eq:massintformula}). Indeed, the particle is still of kind $q_0$, but the contributions to the force of the tails of the ${\cal S}$ are only partial due to the finite size of the system: they decrease in a way that can be dramatically fast. 
\item{\em back to the monovariate regime}, $h_+(q) \leq h$: the ${\cal S}$ shape has rotated in phase-space in such a way that its tails do not influence anymore the motion of its central part. The force is again given by equation (\ref{eq:accvelt2})
and we are ready to proceed until next crossing, where everything will start over, but with a modified ${\cal S}$ shape.
\end{enumerate}
The above description of the dynamical process is correct for $|q| \leq q_{\rm M}/2$. When $|q| \geq q_{\rm M}/2$ (second panel of Fig.~\ref{fig:forcetime}), the phase (iii) disappears, and the end of phase (ii) is increasingly suppressed with $|q|$. Indeed, a regime $h_-(q) \leq h \leq h_{\rm c}(q)$ now appears, during which the tails of the ${\cal S}$ gain energy with time, increasingly with $|q|$, although we did not estimate analytically the transitional value of $q$ that demarcates net positive energy gain from net negative energy gain during an orbital time.

\subsection{Calculation of the corrected motion}

The calculations of the force during the different phases of the motion performed in previous section and in Appendices \ref{app:timesdetails} and \ref{sec:forcecalc} allow us to compute the velocity and the position at third order in $q$. The velocity can be written, for $h \ge h_+(q)$,
\begin{eqnarray}
  v(q,h)& = & v_{\rm b}(q,h)+\sum_{i=1}^{4} \left\{ G_i[q,h_{ i+1}(q)]-G_i[q,h_{i}(q)] \right\}+ \nonumber \\
  & & G_5(q,h)-G_5[q,h_5(q)],
\label{eq:velfinalbig}
\end{eqnarray}
where $h_1(q)=0$ and the four successive event times are given by $h_2(q) = {\hat h}_{\rm c}(q)$, $h_3(q)=h_{\rm c}(q)$, $h_4(q)=h_-(q)$ and $h_5(q)=h_+(q)$, whereas
$G_i$ is a primitive of the force $F$
\begin{equation}
G_i(q,h) =\int F(q,h') \ {\rm d}h',
\end{equation}
of which the explicit expression depends on the phase of the motion, hence the index $i$. 

In an analogous way, the position can be expressed as follows, for $h \geq h_+(q)$, 
\begin{eqnarray}
x(q,h) &=& x_{\rm b}(q,h)+[v(q,h)-v_{\rm b}(q,h)] h +\nonumber \\
          & & \sum_{i=1}^{4}\left[ H_i(q,h_{i+1})-H_i(q,h_i) \right] + \nonumber \\
          & & \sum_{i=1}^4 \left[ G_i(q,h_i) h_i-G_i(q,h_{i+1})h_{i+1} \right] + \nonumber \\
         & & H_5(q,h)-H_5(q,h_{5})+ \nonumber \\
         & & G_5(q,h_5)h_5-G_5(q,h)h,
         \label{eq:myxfinalbig}
\end{eqnarray}
where the $q$ dependence of the $h_i$'s is now implicit, $v(q,h)$ is given by equation (\ref{eq:velfinalbig}), and $H_i$ is a primitive of
$G_i$:
\begin{equation}
H_i(q,h) = \int_h G_i(q,h') \ {\rm d}h'.
\end{equation}

The details concerning the actual calculation of $x(q,h)$ and $v(q,h)$ at third order in $q$ are cumbersome and are deferred to Appendix~\ref{app:cormot}. We now quote and discuss the final results, first in the case with no background, $\omega=0$. This case is particularly interesting because quite relevant in the early stages of the evolution of the full system. In particular, we introduce an interesting toy model neglecting the finite extent of the system ($q_{\rm M} \rightarrow \infty$). Then we present the results for $\omega > 0$.  
\subsubsection{The case with no background: $\omega=0$}
\label{sec:omegaeq0}
For $\omega=0$ and $h \geq h_+(q)$, the final expressions at third order in $q$ for the position and the velocity are
\begin{eqnarray}
x(q,h) & = & (x_{00}+x_{01} h + x_{02} h^2) q + \nonumber \\
          &     & (x_{10}+x_{11} h+ x_{12} h^2) q^3, \label{eq:monxw0} \\
v(q,h) & = & (x_{01}+2 x_{02} h) q +\nonumber \\
          &    & (x_{11} + 2 x_{12} h) q^3,
\end{eqnarray}
with 
\begin{eqnarray}
x_{00} & = &-\frac{ a^2 q_{\rm M}^2 [6 b^2 +c ( c-9 b) q_{\rm M}^2]}{c^2 (b- c q_{\rm M}^2)^2} + \frac{ 6 a^2 b}{c^3} \ln\left( \frac{b}{b-c q_{\rm M}^2} \right), \nonumber \\
& & \\
x_{01} & = & -b +\frac{2 a q_{\rm M}^2(c-3 b)}{c (b-c q_{\rm M}^2)} + \frac{6 a b}{c^2} \ln\left( \frac{b}{b-c q_{\rm M}^2} \right), \\
x_{02} &= & 1, \\
x_{10} &= &  a+\frac{a^2 q_{\rm M}^2\left[ -2 b^2 + b (3 b +c) q_{\rm M}^2+ 2 c q_{\rm M}^4( b-  c q_{\rm M}^2)\right]}{(b-c q_{\rm M}^2)^4}, \nonumber \\
& & \\
x_{11} &= & c-\frac{9 a}{b} +\frac{ 4 a q_{\rm M}^2}{b- c q_{\rm M}^2} + \frac{ a b [ 5 b - 3 (4 b + c) q_{\rm M}^2+ 6 c q_{\rm M}^4]}{3(b-c q_{\rm M}^2)^3}, \nonumber \\
& & \\
x_{21} & = & -1. \label{eq:monx21}
\end{eqnarray}
Note that we do not assume that $q_{\rm M}$ is a small parameter here, hence the complexity of the expressions, although these later are meaningful only if $q_{\rm M} < q_{\rm v}$,
with
\begin{equation}
q_{\rm v} \equiv \sqrt{\frac{b}{c}}.
\label{eq:qv}
\end{equation}
The reason for not supposing $q_{\rm M}$ small  is that this parameter corresponds, in phase-space, to the most remote part of the ${\cal S}$ and terms containing it have to be fully taken into account for best accuracy. Of course, a small value of $q_{\rm M}$ will definitely improve the quality of our description: we shall see in next section that there are stringent conditions on how big can $q_{\rm M}$ be for our calculation to be actually valid, when iterating from crossing to crossing. Indeed, for the true dynamical system, the curve $[x(q,t),v(q,t)]$ in general is not a third order polynomial in $q$. This means that the actual remote tails of the ${\cal S}$ shape are poorly approximated by the curve $[x_{\rm b}(q,t),v_{\rm b}(q,t)]$, so the quality level of our force calculation using this approximation is increasingly poor when $q_{\rm M}$ augments: this is a {\em global} effect that affects as well the dynamics of the center of the system. In particular, it will lead, as we shall see in \S~\ref{sec:numexp} when comparing analytical predictions to measurements in numerical experiments, to estimates of the successive crossing times slightly offset from their actual values.  Note of course that this discussion also applies to the case $\omega > 0$ treated below.

While keeping aware of these limitations,  it is still interesting to see what happens when one relaxes the condition $q_{\rm M} < q_{\rm v}$ and assumes that $q_{\rm M}$ is arbitrarily large.  This means that the equation $x(q,h)=x(q_0,h)$ will always have either one or three solutions (including the trivial one, $q_0$), which simplifies considerably the calculations of the equations of motion, which no longer require taking care of the extension of the system up to $q_{\rm M}$: we only have to account for the three first phases, (i), (ii) and (iii), of \S~\ref{sec:forcecalc}.  In other words, only terms containing $F_1$, $F_2$ and $F_3$ contribute to the dynamics.  The expressions for the position and the velocity are then rather simple, for $h \geq h_{\rm c}(q)$,
\begin{eqnarray}
x(q,h) & = & \left[ -\left(\frac{6 a b}{c^2}+b \right) h+\left(2 -  \frac{3b}{c} \right) h^2+\right. \nonumber \\ & &\quad \quad \left. \frac{6 a b}{c^2} \left( h+\frac{a}{c} \right) \ln \left( 1+ \frac{c h}{a} \right) \right] q+ \nonumber\\
& & \left[ a +\left( c-\frac{9 a}{b} \right) h+ h^2 \right] q^3, \label{eq:xnoqmax} \\
v(q,h) & = & \left[-b +\left(4- \frac{6b}{c} \right) h +\frac{6 a b}{c^2} \ln\left( 1+\frac{ c h}{a} \right) \right] q +\nonumber \\
& & \left( c-\frac{9a}{b} + 2 h\right) q^3. \label{eq:vnoqmax}
\end{eqnarray}
Interestingly enough, we shall see by comparison with numerical experiments that this simplification of the dynamics provides reasonable results, at least from the qualitative point of view. The main reason for this is that the point $[x_{\rm b}(\pm q_{\rm v},h),v_{\rm b}(\pm q_{\rm v},h)=0]$ is a stationary one. In realistic situations, only the points with $|q| \leq q_{\rm v}$ are expected to contribute, in practice, to the dynamics. However, $q=q_{\rm v}$ is a rather ``large'' value of $q$, where one expect the approximation of the curve $[x(q,h),v(q,h)]$ by a third order polynomial in $q$ to be poor, which explains why equations (\ref{eq:xnoqmax}) and (\ref{eq:vnoqmax}) should be seen as a toy model rather than a fully realistic description of the equations of motion.   

\subsubsection{Taking into account the background: $\omega > 0$}
The case $\omega \neq 0$ is more complex, because the analog of the expression for $x_{00}$ in equation (\ref{eq:monxw0}) is not fully analytical. However, except for one integral $Y$ that has to be estimated numerically, the full expressions for the position and the velocity can still be explicitly written at third order in $q$:\footnote{A {\tt Mathematica} notebook is available on request from the author.}
\begin{eqnarray}
x(q,h) & = & (x_{00}+x_{01} h + x_{02} h^2-\frac{b}{\omega} \sin \omega h) q + \nonumber \\
          &     & (x_{10}+x_{11} h+ x_{12} h^2 + \nonumber \\
          &     & a \cos \omega h + \frac{c}{\omega} \sin \omega h ) q^3, \label{eq:monxw} \\
v(q,h) & = & (x_{01}+2 x_{02} h-b \cos \omega h) q +\nonumber \\
          &    & (x_{11} + 2 x_{12} h + c \cos \omega h- a \omega \sin \omega h) q^3,
\end{eqnarray}
with
\begin{eqnarray}
x_{00} & = & Y+ \frac{3 b c - c^2 - a^2 \omega^2}{\omega^2 (c^2+a^2 \omega^2)} \arccos\left( \frac{b-c q_{\rm M}^2}{\sqrt{T}} \right)^2 - \nonumber \\
      & & \frac{6 a b}{\omega (c^2+a^2 \omega^2)} \arccos\left( \frac{b-c q_{\rm M}^2}{\sqrt{T}}\right) \ln\left( \frac{a b \omega}{\sqrt{T}}\right) , \\
x_{01} & = & \frac{2(-3 b c + c^2 + a^2 \omega^2) }{\omega (c^2 + a^2 \omega^2)}\arccos\left( \frac{b- c q_{\rm M}^2}{\sqrt{T}}\right) + \nonumber \\
         & & \frac{6 a b }{c^2 + a^2 \omega^2} \ln\left( \frac{b}{\sqrt{T}} \right),\\
x_{02} &= & 1, \\
x_{10} & = &  
         \frac{ a^2 b^2 q_{\rm M}^2 ( 3 q_{\rm M}^2-1)}{3 T^2}  + \nonumber \\
          &  & \frac{a b}{3  \omega T^2} \left[ 2 b c q_{\rm M}^2(4-9 q_{\rm M}^2) + b^2 (12 q_{\rm M}^2 -5) + \right. \nonumber \\
           & &  \left. 3 q_{\rm M}^4 (2 q_{\rm M}^2-1)(c^2 + a^2 \omega^2) \right] \arccos\left( \frac{b - c q_{\rm M}^2}{\sqrt{T}}\right)+\nonumber \\
          & & \frac{2}{\omega^2} \arccos\left( \frac{b - c q_{\rm M}^2}{\sqrt{T}}\right)^2, \\
x_{11} & = & -\frac{9 a }{b} - \frac{a b}{3 T^2}\left[ 2 b c q_{\rm M}^2 ( 4 -9 q_{\rm M}^2)
+ b^2 ( 12 q_{\rm M}^2 - 5) + \right. \nonumber \\ 
         &  & \left.  3 q_{\rm M}^4 ( 2 q_{\rm M}^2 -1) (c^2 + a^2 \omega^2) \right]+ \nonumber \\ 
 & & \frac{4}{\omega} \arccos\left( \frac{b- c q_{\rm M}^2}{\sqrt{T}}\right), \\
x_{12} & = & -1, \label{eq:monx12w}
\end{eqnarray}
where $T$ and $Y$ are given by equations (\ref{eq:eqforT}) and (\ref{eq:eqforY}), respectively.

Note that the arbitrarily large $q_{\rm M}$ assumption cannot be used to simplify the calculations and to construct a toy model similarly as in the case $\omega=0$. Indeed, with $\omega \neq 0$, the system is rotating in phase-space due to the harmonic motion. Therefore, the cut at $q_{\rm M}$ becomes a fundamental part of the dynamics: without this cut-off, a singular, non physical behavior would appear at a time $h=\arccos(-c \sqrt{c^2 + a^2 \omega^2})/\omega$, where $q_{\rm c}(h)$ diverges (equation~\ref{eq:qcexpr}). 

However, one can still display, for completeness, the large $\omega$ regime:
\begin{eqnarray}
x(q,h) & \simeq & \left[ h^2 +\frac{\pi}{\omega} h - \frac{b}{\omega} \sin(\omega h) \right]q +\nonumber \\
 &  & \left[ -h^2 +\left(\frac{2 \pi}{\omega}-\frac{9 a }{b} \right) h + \right.  \nonumber \\
 &  &  \quad \quad \quad \left. a \cos( \omega h) +\frac{ c}{ \omega }\sin(\omega h)\right] q^3, \\
v(q,h) & \simeq &  \left[ 2 h + \frac{\pi}{\omega} - b \cos(\omega h) \right] q +\nonumber \\
 & & \left[-2h+\frac{2\pi}{\omega}- \frac{9 a}{b} +  c \cos(\omega h) - a \omega \sin(\omega h)\right]q^3. \nonumber \\
& & 
\end{eqnarray}
\subsection{State of the system at next crossing time: a recurrence formula}
\label{sec:recufor}
The expressions calculated in previous section are of the form
\begin{equation}
x(q,h)  =  {\cal E}(h,a,b,c,q_{\rm M},\omega) q - {\cal A}(h,a,b,c,q_{\rm M},\omega) q^3, 
\end{equation}
\begin{equation}
v(q,h)  =   {\cal B}(h,a,b,c,q_{\rm M},\omega) q - {\cal C}(h,a,b,c,q_{\rm M},\omega)q^3.
\end{equation}
with, naturally,  ${\cal B} = \partial {\cal E}/\partial h$ and ${\cal C} = \partial {\cal A}/\partial h$.
Crossing time corresponds to the smallest possible strictly positive value of $h$ with
\begin{equation}
{\cal E}(h,a,b,c,q_{\rm M},\omega) = 0.
\label{eq:collapseteq}
\end{equation}
Note that this equation has an analytic solution in the limit $\omega \rightarrow 0$ (that we do not write here, for simplicity), since in this case
${\cal E}$ is a second order polynomial in $h$ (equation \ref{eq:monxw0}). 

\begin{enumerate}
\item{\it Recurrence formula:} resolving equation (\ref{eq:collapseteq}) (explicitly if $\omega=0$, numerically otherwise) allows us to rewrite the position and the velocity at crossing time in the same form as in equations (\ref{eq:crossingx}) and (\ref{eq:crossingy}):
\begin{eqnarray}
x(q,t_{{\rm c},n})&=& a_n q^3, \label{eq:recux} \\
v(q,t_{{\rm c},n}) & = & -b_n q + c_n q^3, \label{eq:recuv}
\end{eqnarray}
and to establish a recurrence of the state of the system at successive crossing times as a function of crossing time number $n$:
\begin{eqnarray}
a_{n+1} &=&{\cal A}(h_n,a_n,b_n,q_{{\rm M},n},\omega_n), \\
b_{n+1} &=&{\cal B}(h_n,a_n,b_n,q_{{\rm M},n},\omega_n), \\
c_{n+1} &=&{\cal C}(h_n,a_n,b_n,q_{{\rm M},n},\omega_n), \\
t_{{\rm c},n+1} &= & t_{{\rm c},n}+h_n, \\
h_n &=&\min_{h >0}\left\{h \, |\, {\cal E}(h,a_n,b_n,c_n,q_{{\rm M},n},\omega_n)=0  \right\},
\end{eqnarray}
where a transformation $q \rightarrow -q$ has been performed  to take into account of the rotation of the system in phase-space by half a circle. This transformation is made so to preserve, in general, the positive nature of the coefficients $a_n$, $b_n$ and $c_n$. 

\item{\it Choice of the extension of the system, $q_{{\rm M},n}$:} at each iteration, a crucial step consists in defining the Lagrangian size $q_{\rm M}=q_{{\rm M},n}$ of the system between crossing times $n$ and $n+1$. A natural choice for such a scale is to take the point where velocity reaches is extremum,
\begin{equation}
\frac{\partial v}{\partial q_{{\rm M},n} }\equiv 0,
\end{equation}
i.e.
\begin{equation}
q_{{\rm M},n} \equiv \sqrt{\frac{b_n}{3c_n}},
\label{eq:defiqn}
\end{equation}
which will be our choice in the subsequent analyses. 

\item{\it Recurrence for the background:} we may assume that $q_{{\rm M},n}$ is a decreasing function of $n$, to reflect the fact that at each crossing time, the central ${\cal S}$ shape loses some material from its two tails, and that this material contributes to the background:
\begin{equation}
\omega_n = \sqrt{2 \rho_n},
\end{equation}
where $\rho_n$ is given by the recursion 
\begin{eqnarray}
\rho_{n+1} &=& \rho_n + \Delta \rho_n, \label{eq:rhon} \\
\Delta \rho_n &=& \beta \frac{M(q_{{\rm M},n})-M(q_{{\rm M},n+1})}{2 x(q_{{\rm M},n},t_{{\rm c},n+1})},
\label{eq:recuro}
\end{eqnarray}
and $\beta={\cal O}(1)$ is a parameter that specifies the typical extension of the region filled up by the matter escaping from the tails of the ${\cal S}$, while, with our initial set-up, $\rho_0 \equiv 0$. 
 In reality, the system looks like a spiral in phase-space, which means that the two parts of the tails contributing to the background are elements of this spiral. These elements are getting elongated with time as a result of the differential rotation speed of the system in phase-space. They occupy a ring of approximately fixed boundaries, due to the (supposedly) nearly adiabatic nature of the halo, and progressively fill this ring up in an homogeneous way. Combining arguments based on total energy conservation and stationarity would probably allow one to compute iteratively the shape of this ring rather accurately but this would go beyond the scope of this paper. Instead, we chose to keep $\beta$ as a single but adjustable parameter when performing comparisons with numerical experiments. However, it might, in reality, depend on time, i.e. on $n$. Choosing it fixed implicitly assumes some kind of similarity behavior, which is not guaranteed. 

\item{\it Condition of validity:} 
for the recurrence to make sense
it is required that $x(q_{{\rm M},n},t_{{\rm c},n})$ and $v(q_{{\rm M},n},t_{{\rm c},n})$ have a chance to remain close enough to their actual values. We already discussed partly this issue in \S~\ref{sec:omegaeq0}, where we argued that the analytical predictions are expected to behave better when $q_{\rm M}$ is kept smaller.
In fact, because the equations of motions are integrated for $h \geq h_+(q) \geq h_{\rm c}(q)$, it is needed, for having a chance to have a realistic description of the system at $q_{{\rm M},n}$, to have
\begin{equation}
h_n  \ga \max[ h_{\rm c}(q_{{\rm M},n+1}), h_+(q_{{\rm M},n+1}) ], 
\label{eq:condival}
\end{equation}
where $h_{\rm c}$ and $h_+$ are given by their {\em approximations}, equations~(\ref{eq:hcap}), (\ref{eq:hpmexp}) or (\ref{eq:hpmexpw0}) (hence, one can have $h_{\rm c} > h_+$, as illustrated by second panel of Fig.~\ref{fig:charactimes}). 
If this non trivial (necessary, but not sufficient) condition is not fulfilled, it means that our recurrence will definitely go away from a good approximation of the true dynamics. 
\end{enumerate}
\subsection{Discussion}
\begin{figure}
\centerline{\hbox{
\psfig{file=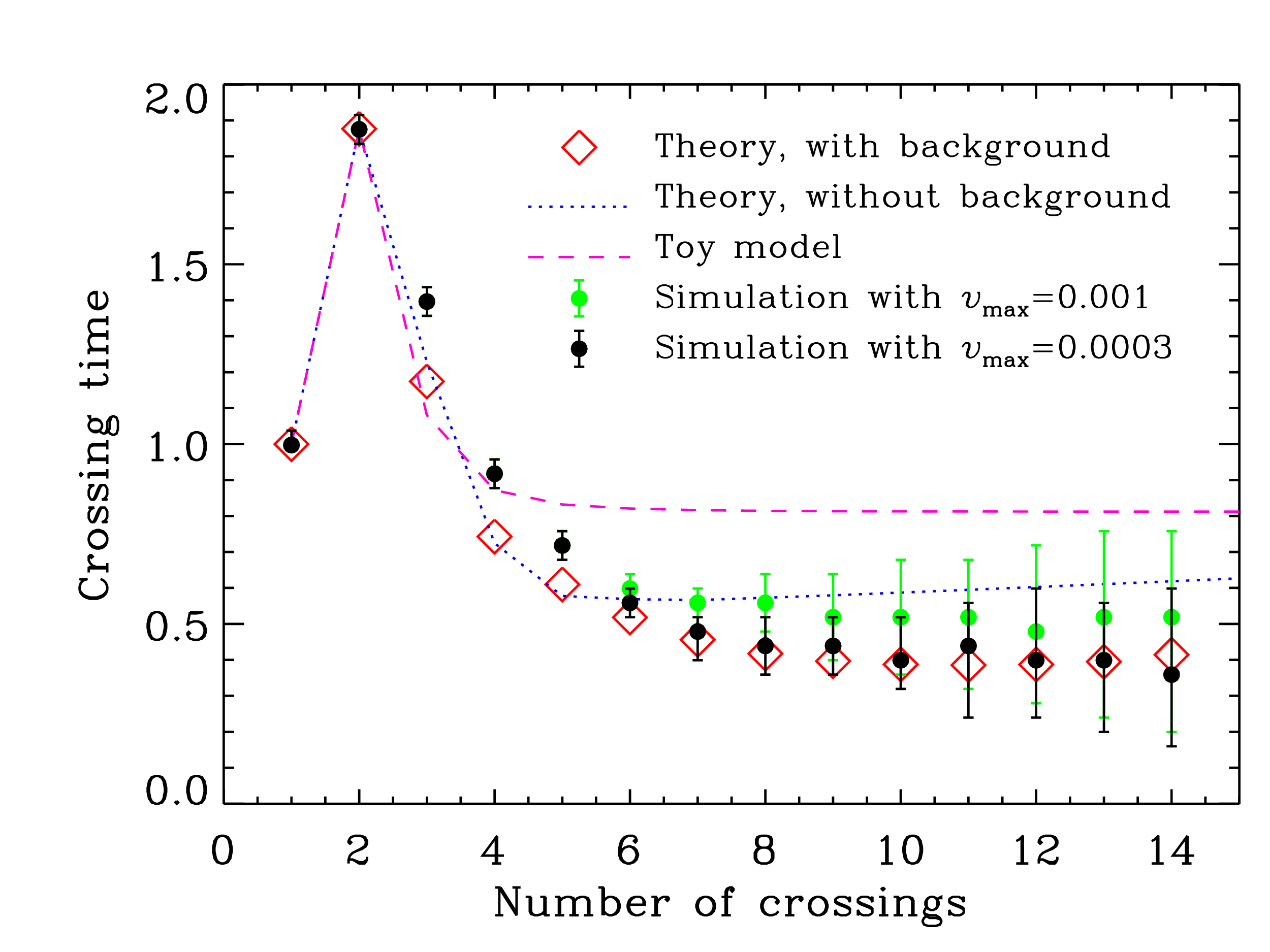,width=8.6cm}}}
\centerline{\hbox{
\psfig{file=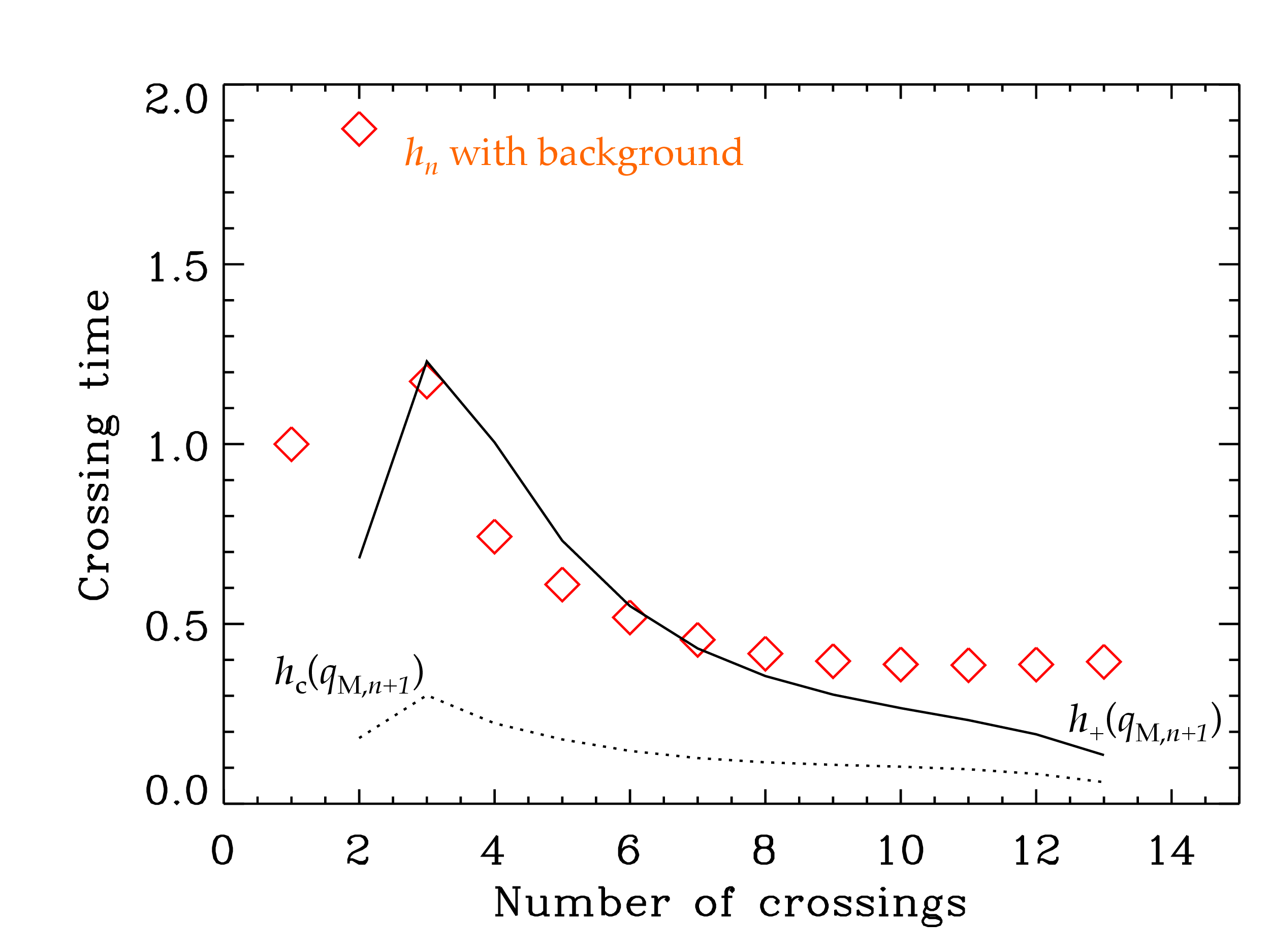,width=8.6cm}}}
\caption[]{The crossing time, $t_{{\rm c},n}$, as a function of crossing number, $n$. In the {\em top panel}, the three models (red symbols: with background, dotted blue: without background; dashed pink: toy model) are compared to the simulations of \S~\ref{sec:numexp} (black and green circles). The errorbars on the simulation points reflect the uncertainty induced by the gap in time between successive snapshots, combined with human error: the successive crossing times have been determined by visual inspection, which is the main source of fluctuations at late times. The {\em bottom panel} compares the crossing time obtained with background to $h_{\rm c}(q_{{\rm M},n+1})$ (dots) and $h_+(q_{{\rm M},n+1})$ (solid) to test the validity of the model (equation \ref{eq:condival}). Of course, there is no dotted nor solid line at $n=1$, where Zel'dovich dynamics provides an exact answer.}
\label{fig:crossingtime}
\end{figure}

We now examine what is the outcome of the recurrence with the initial conditions of \S~\ref{sec:zeldo}. Figure \ref{fig:coefficients} shows the evolution of the coefficients $a_n$, $b_n$ and $c_n$ of equations (\ref{eq:recux}) and (\ref{eq:recuv}) as functions of number of crossings. 
Three cases are shown on the figure: a full calculation with an additional increasing halo background consistent with the mass loss of the ${\cal S}$ in the tails, the calculation ignoring the effect of the background ($\omega_n=0$) and the toy model (still corresponding to $\omega_n=0$) described in \S~\ref{sec:omegaeq0}. Several important properties show up:
\begin{figure}
\centerline{\hbox{
\psfig{file=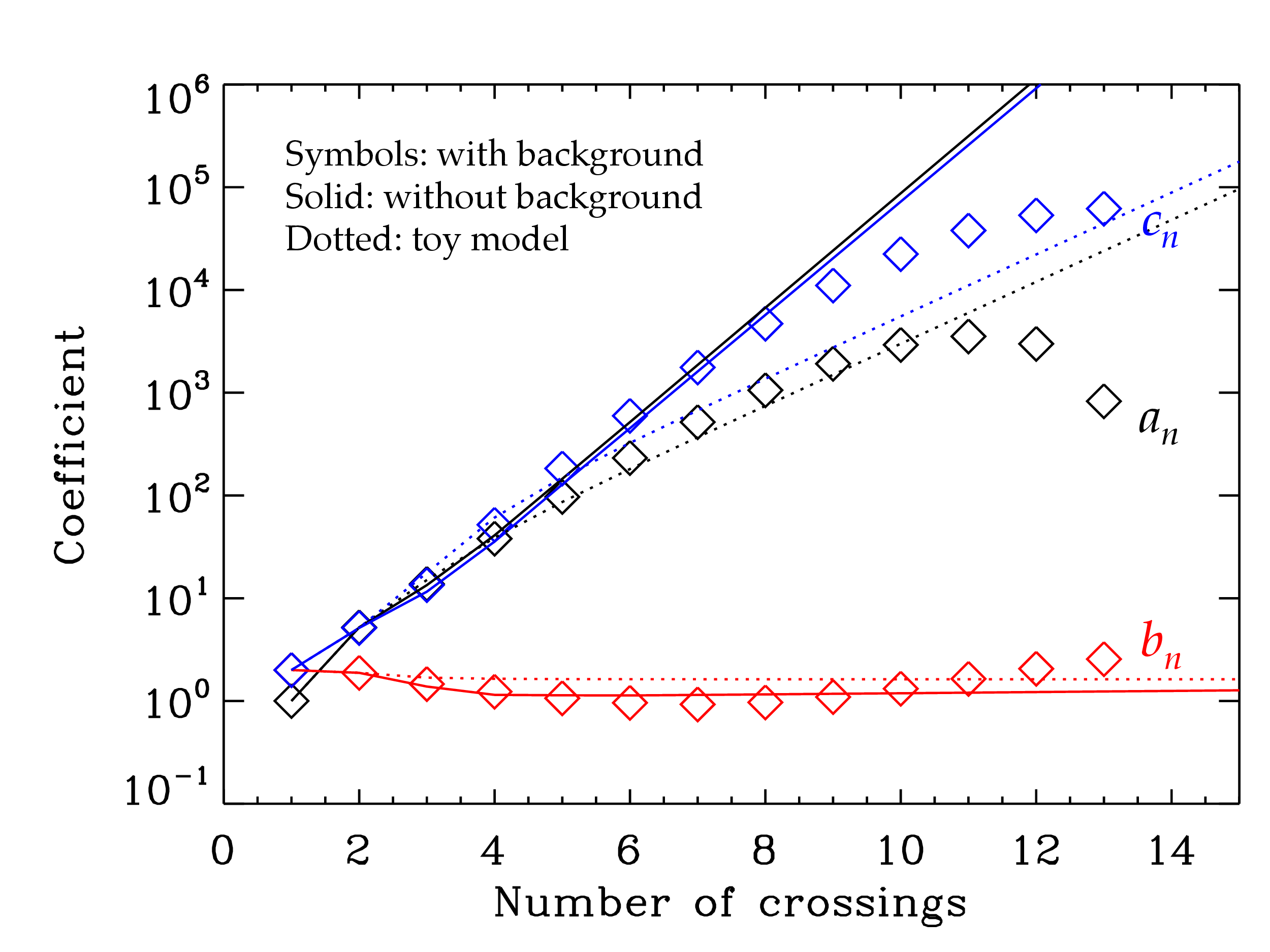,width=8.6cm}
}}
\caption[]{The coefficients $a_n$ (black), $b_n$ (red) and $c_n$ (blue) describing the shape of the central ${\cal S}$ shape at crossing times (equations \ref{eq:recux} and \ref{eq:recuv}) as functions of the number of crossings. The symbols correspond to the full calculation with $\omega_n >0$ (equations \ref{eq:monxw} to \ref{eq:monx12w}) and $\beta=1.5$ in equation (\ref{eq:recuro}). This value of $\beta$ was chosen 
such that the mass interior to $q_{{\rm M},n}$ obtained at successive crossing times roughly fits the one measured in the simulation with $v_{\rm max}=0.0003$ of \S~\ref{sec:numexp}. The solid lines correspond to the case $\omega=0$ (equations \ref{eq:monxw0} to \ref{eq:monx21}) and the dots to the toy model (still assuming $\omega=0$) given by equations (\ref{eq:xnoqmax}) and (\ref{eq:vnoqmax}).}
\label{fig:coefficients}
\end{figure}
\begin{description}
\item[(a)] the three models match rather well each other up to fourth crossing. The effect of the halo background on the dynamics of the center of the system becomes critical only when $n \geq 5$ which corresponds to a rather advanced time in the evolution of the system;
\item[(b)] the velocity scale coefficient $b_n$ remains approximately constant with time and of order of unity, whatever model.
\item[(c)] the coefficients $a_n$ and $c_n$ augment approximately geometrically with number of crossings, for $n \ga 4$. In the case $\omega=0$, this regime extends indefinitely, and the geometric amplitude is the same for $a_n$ and $c_n$, but different for the toy model and the ``realistic'' calculations. In the presence of background, on the other hand, the geometric amplitude is different for $a_n$ and $c_n$ and this exponential behavior is limited in duration. 
\item[(d)] For $\omega >0$, the geometric growth of $a_n$ and $c_n$ gets indeed interrupted at late times, $n \sim 9$, where $a_n$ and $c_n$ start bending. Furthermore, the symbols on Fig.~\ref{fig:coefficients} stop at large time because $a_n$ changes sign at the 14th crossing, which corresponds to a ``reversal'' of the shape of the ${\cal S}$. Whether this instability in our analytical model reflects something inherent to the true dynamical system remains to be understood. As discussed in subsequent section, \S~\ref{sec:numexp}, we unfortunately do not have at our disposal a sufficiently cold set up to demonstrate the existence of this instability with numerical experiments. 
\end{description}
For completeness, top panel of Fig.~\ref{fig:crossingtime} displays the crossing time as a function of number of crossings for the three models of the dynamics. After a peak corresponding to what could be described as a violent relaxation period, the system reaches a plateau. Bottom panel of Fig.~\ref{fig:crossingtime} is a test for the model with background of the validity condition (\ref{eq:condival}), which is verified, except for $3 \leq n \leq 6$ where a marginal violation can be observed. 
\begin{figure}
\centerline{\hbox{
\psfig{file=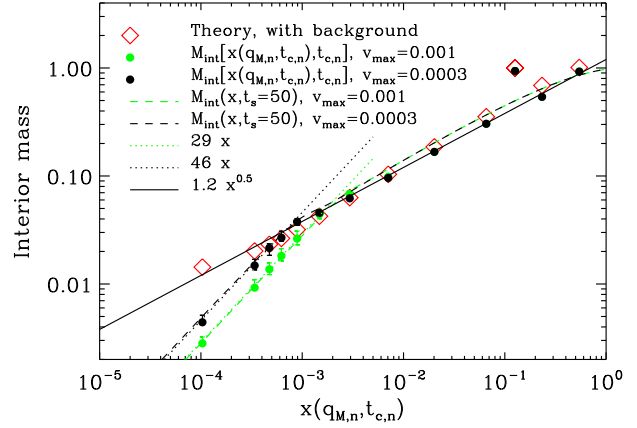,width=8.9cm}
}}
\caption[]{The total mass interior to $x(q_{{\rm M},n},t_{{\rm c},n})$ obtained at different crossing times, where $q_{{\rm M},n}$ corresponds to the Lagrangian coordinate of the extrema of the theoretical velocity and $t_{{\rm c},n}$ is the theoretical $n^{\rm th}$ crossing time. Therefore the $x$ axis of the plot is entirely determined by the theory, because we do not have a simple mean to estimate accurately the value of $x(q_{{\rm M},n},t_{{\rm c},n})$ from the simulation, except at early times. This is due to the fact that opposite borders of the waterbag ``slide'' with respect to each other. The symbols with errorbars represent measurements in the two simulations described in \S~\ref{sec:numexp} (black and green respectively for $v_{\rm max}=0.0003$ and $v_{\rm max}=0.001$), which can be directly compared to the theoretical prediction including the contribution of the halo background (red symbols). The free parameter $\beta$ in equation (\ref{eq:recuro}) has been tuned so that theory globally matches the simulations in their range of validity. This range is limited by the warm nature of the waterbag set ups, inducing the presence of a small core with $M_{\rm int}(x) \propto x$ when $x$ is very small (green and black dotted lines). If the system was stationary, the mass $M_{{\rm int},n}$ interior to $x(q_{{\rm M},n},t_{{\rm c},n})$ at crossing time $n$ should be equal to the interior mass measured at the same position, but at lates times. This latter is shown as a function of position for both simulations (black and green dashes), and indeed approximately matches $M_{{\rm int},n}$, except, not surprisingly, at collapse time ($n=1$). Finally, the solid line corresponds to the power-law behavior $M_{\rm int}(x) \propto x^{1/2}$ conjectured by \cite{Binney}.}
\label{fig:massprofile}
\end{figure}

It is interesting at this point to discuss the properties of the system assuming, following point (c) above, that $a_n$ and $c_n$ vary geometrically with time, as well as $b_n$: 
\begin{equation}
a_n \propto \gamma_a^n, \quad b_n \propto \gamma_b^n, \quad c_n \propto \gamma_c^n. \label{eq:geomprop}
\end{equation}
According to point (b) above, we should have $\gamma_b \simeq 1 < \gamma_c$, which is clearly {\em incompatible with self-similarity in phase-space} of the center of the system, that would imply $\gamma_b \simeq \gamma_c$. This deviation from self-similarity might be due to limitations of our theoretical model. On the other hand, as discussed further in \S~\ref{sec:numexp}, we shall see that measurements in numerical simulations suggest that we should have $\gamma_b \simeq \gamma_a^{-1/4} \la 1$, hence $\gamma_b$ still smaller than $\gamma_c$. 

We now try to compute, in the framework of the recurrence proposed in \S~\ref{sec:recufor}, the properties of the expected projected density profile, $\rho(x)$, assuming property (\ref{eq:geomprop}) and that the halo is stationary.  With these hypotheses, the projected density at position
$x(q_{{\rm M},n},t_{{\rm c},n})$ reads
\begin{equation}
\rho[x(q_{{\rm M},n},t_{{\rm c},n})]  \simeq \frac{M(q_{{\rm M},n})}{2 x(q_{{\rm M},n},t_{{\rm c},n})}+\sum_{m < n} \Delta \rho_m.
\end{equation}
From equations (\ref{eq:geomprop}) and (\ref{eq:defiqn}) we have
\begin{equation}
q_{{\rm M},n} \propto \left( \frac{\gamma_b}{\gamma_c} \right)^{n/2},
\end{equation}
hence (since $q_{{\rm M},n}$ is small),
\begin{equation}
M(q_{{\rm M},n}) \propto \left( \frac{\gamma_b}{\gamma_c} \right)^{n/2}, \quad x(q_{{\rm M},n},t_{{\rm c},n}) \simeq {\hat x}\, 
\gamma_x^n , 
\end{equation}
with 
\begin{equation}
\gamma_x \equiv \frac{\gamma_a \gamma_b^{3/2}}{\gamma_c^{3/2}} < 1
\end{equation}
and where ${\hat x}$ is some number. Note as well that $x(q_{{\rm M},n},t_{{\rm c},n-1}) \propto \gamma_x^n$.  The property $\gamma_x < 1$ steams from the fact that $x(q_{{\rm M},n},t_{{\rm c},n})$ should strictly decrease with $n$ (unless the harmonic contribution completely dominates). Therefore, in equations (\ref{eq:rhon}) and (\ref{eq:recuro}),
\begin{equation}
\Delta \rho_n \simeq  {\hat \rho} \gamma_{\rho}^n,
\end{equation}
where ${\hat \rho}$ is some number and
\begin{equation}
\gamma_\rho \equiv \frac{\gamma_c}{\gamma_a \gamma_b}.
\end{equation}
As a result, we have
\begin{eqnarray}
\rho( {\hat x} \gamma_x^n ) = & {\tilde \rho} \gamma_\rho^{n}+{\hat \rho} \frac{ \gamma_\rho^{n}-1}{\gamma_\rho-1}, \quad & \gamma_\rho \neq 1, \\
                                          = & {\tilde \rho}+(n-1) {\hat \rho}, &\gamma_\rho = 1,
\end{eqnarray}
where ${\tilde \rho}$ is some number. So, for $x \ll 1$ (large $n$),
\begin{eqnarray}
\rho(x) \propto & x^\gamma, & \gamma_\rho > 1, \\
  \propto & \ln x,  & \gamma_\rho=1, \label{eq:rholog} \\
  = & {\rm constant}, & \gamma_\rho < 1,
\end{eqnarray}
with
\begin{equation}
 \gamma\equiv {\frac{\ln\gamma_\rho}{\ln\gamma_x}} < 0.
\end{equation}
This means that for the toy model as well as for the case when the harmonic background is neglected, $\gamma_c \simeq \gamma_a$ and $\gamma_b \simeq 1$, one expects a logarithmic divergence of the projected density at small $x$. If the harmonic contribution is taken into account, we notice from Fig.~\ref{fig:coefficients} that  $\gamma_c > \gamma_a$ for $n \ga 4$ while $\gamma_b \simeq 1$, supporting at least in some scaling range a power-law singularity paradigm for the projected density, as illustrated by Fig.~\ref{fig:massprofile}. Then, if our calculations are qualitatively valid at large crossing times, the instability observed at late time for $\omega > 0$ [item (d) above] suggests the existence of a very small core, but this is very speculative.

\section{Comparison with controlled numerical experiments}
\label{sec:compnum}
\label{sec:numexp}
We analyse two waterbag simulations performed with the Vlasov solver presented and tested in \cite{CT08,CT12}. This latter exploits directly Liouville theorem to follow accurately, in an entropy conserving fashion, the borders of a patch of constant phase-space density --the waterbag-- with a self-adaptive, orientated polygon. The waterbag, of total mass $m=1$, initially occupies in phase-space the ellipse of equation
\begin{equation}
(x/x_{\rm max})^2+(v/v_{\rm max})^2=1,
\label{eq:ellipse}
\end{equation}
with $x_{\rm max}=1$, $v_{\rm max}=0.001$ for the first simulation and $v_{\rm max}=0.0003$ for the second one, which corresponds in both cases to a rather cold set-up, as needed to test our analytical predictions.  Inside the ellipse defined by equation (\ref{eq:ellipse}), the phase-space density reads
\begin{equation}
f(x,v)=\frac{m}{\pi x_{\rm max} v_{\rm max}}.
\end{equation}
The initial projected density does not depend on the initial velocity dispersion and reads
\begin{equation}
\rho_{\rm i}(x)=\frac{2 m}{\pi x_{\rm max}} \sqrt{1-(x/x_{\rm max})^2}
\simeq {\bar \rho}_0(1-3 \alpha x^2),
\end{equation}
with
\begin{equation}
{\bar \rho}_0=\frac{2 m}{\pi x_{\rm max}}, \quad \alpha=\frac{1}{6 x_{\rm max}^2}.
\end{equation}
To match the mass of our theoretical model and that of the simulated system, 
one needs to set the initial value of $q_{\rm M}$ to $q_{\rm M}=0.3722 < 1/\sqrt{3}$. This is obtained
by solving the equation
\begin{equation}
m=2 {\bar \rho}_0\left[ \frac{q_{\rm M}}{\sqrt{\alpha}}  -\alpha \left( \frac{q_{\rm M}}{\sqrt{\alpha}} \right)^3 \right].
\end{equation}

\begin{figure*}
\centerline{\hbox{
\psfig{file=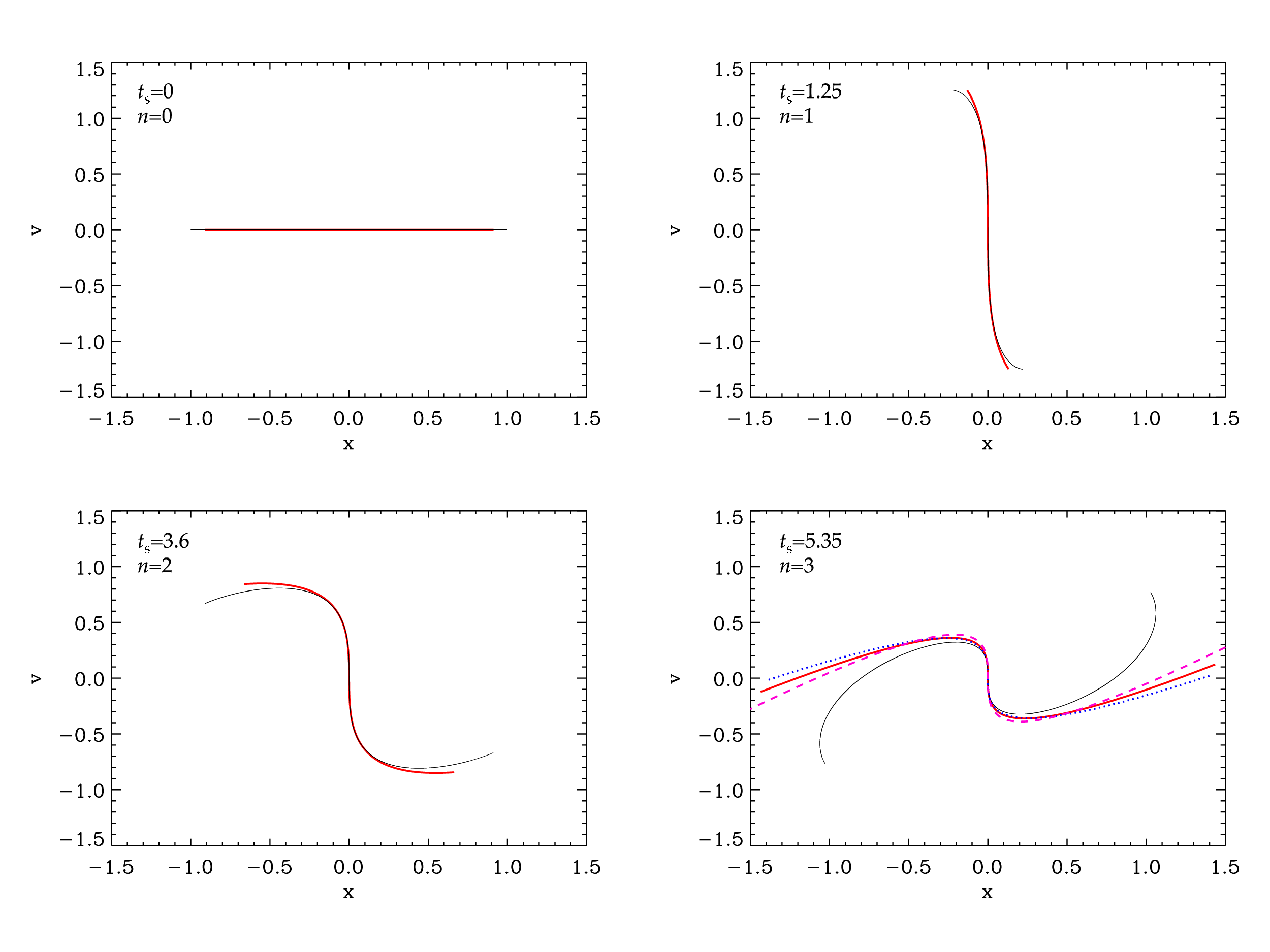,width=16cm}
}}
\caption[]{Evolution in phase space of the distribution function at early times. Each panel represents a state of the evolution of the system at a given crossing time (except for the upper-left one which corresponds to initial conditions), with the simulation time $t_{\rm s}$ indicated on the upper left part, in addition to the number of crossings, $n$. On each panel, there is a black curve, which corresponds to the simulated $f(x,v)$ with $v_{\rm max}=0.0003$ in equation (\ref{eq:ellipse}). Note that this is not really a curve but the contour of a single waterbag. However the very cold nature of the system does not allow us here to really distinguish the borders of the waterbag: for instance, the initial ellipse looks like a flat line. The red solid curve corresponds to our analytical model, that includes a harmonic contribution when $n \geq 3$, with $\beta=1.5$ in equation (\ref{eq:recuro}). The blue dotted curve on the lower-right panel is the same as the red one, but the harmonic background is neglected in the dynamics ($\omega=0$). Finally, the pink dashed curve on the lower-right panel corresponds to the toy model for the $\omega=0$ case, when used from second crossing time. On the bottom-right panel, the tails of the theoretical curves are truncated at $q=q_{{\rm M},2}$. Note furthermore, on this panel, that a time shift has been applied to the theory in order to synchronize it with the simulation, due to the difference between the theoretical and the numerical crossing time.}
\label{fig:evolvpha}
\end{figure*}

\begin{figure*}
\vskip -0.5cm
\centerline{\hbox{
\psfig{file=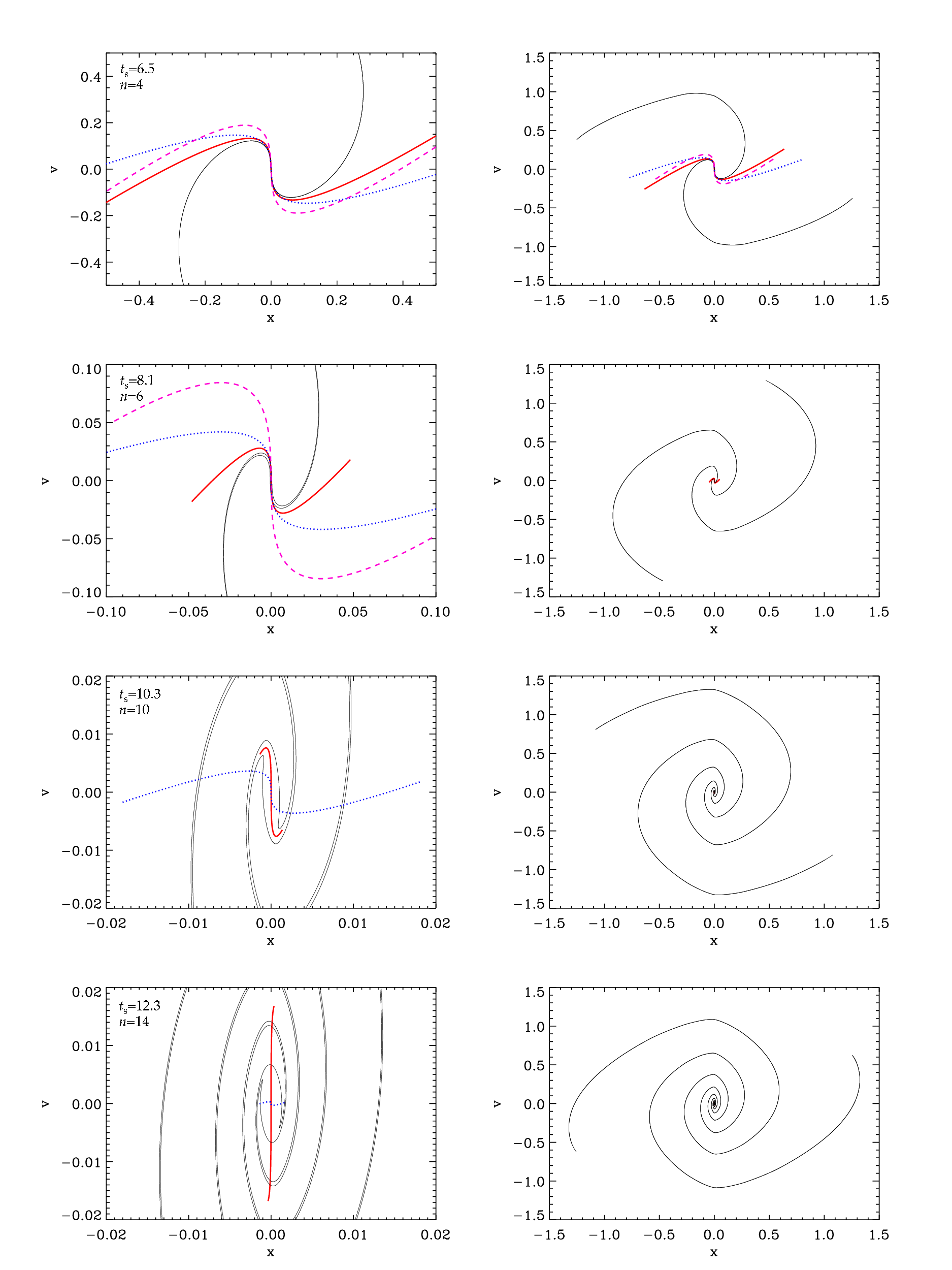,width=16cm}
}}
\caption[]{Evolution in phase space of the distribution function at later times. This figure is the follow-up of Fig.~\ref{fig:evolvpha} but the presentation is slightly different. Each line of panels corresponds to a state of the evolution of the system at a given crossing time, with a zoom on the center of the system on left panel, while a global view of the phase-space distribution function is presented on the right panel. The curves are the same as in bottom-right panel of Fig.~\ref{fig:evolvpha}. When $n \geq 6$, only the red curves are represented on right panels for clarity. The toy model is only displayed up to $n=6$, as it clearly diverges from the true solution.  Again, similarly as in bottom-right panel of Fig.~\ref{fig:evolvpha}, a time shift is applied to theory in order to synchronize it with the simulation to account for the difference between the theoretical and the measured crossing times. At late times, the appearance of an elliptic core due to the slightly warm nature of the waterbag simulation makes the comparison between theory and measurements somewhat irrelevant. On the bottom-left panel, the tails of the red curve have a curvature opposite to previous times. This is due to a sign change of the coefficient $a_n$ in equation (\ref{eq:recux}).}
\label{fig:evolvphaad}
\end{figure*}

Figures~\ref{fig:evolvpha} and \ref{fig:evolvphaad} show the evolution of the phase-space distribution function at successive crossing times for the simulation with $v_{\rm max}=0.0003$ and compare it to our theoretical predictions for the shape of the central ${\cal S}$.  

Up to first crossing time (top-right panel of Fig.~\ref{fig:evolvpha}), the Zel'dovich dynamics of \S~\ref{sec:zeldo} (equations \ref{eq:xdyn} and \ref{eq:vdyn}) describes exactly the evolution of the system and the difference between analytical prediction (red curve) and simulation (black curve) is just due to the fact that the theoretical system is initially slightly different from the simulated one, increasingly far away in the tails. 

The description of the transition between first crossing time and second one represents the first non trivial result of our theoretical investigations, in the $\omega=0$ case. The red curve on bottom-left panel of Fig.~\ref{fig:evolvpha} is given by equations (\ref{eq:monxw0}) to (\ref{eq:monx21}), when starting from the state at collapse time given by the red curve on the top-right panel. The match between the red curve and the black one is very good, especially if one remembers that theory is valid only to third order in $q$ and is therefore expected to be very approximate in the tails of the ${\cal S}$. 

In bottom-right panel of Fig.~\ref{fig:evolvpha}, that corresponds to third crossing time, we added the presence of a background with $\beta=1.5$ in equation (\ref{eq:recuro}). The red curve is thus now given by equations (\ref{eq:monxw}) to (\ref{eq:monx12w}), starting from the state at second crossing given by the red curve on bottom-left panel.  The value of $\beta$ was adjusted manually to visually reproduce as well as possible the mass interior to $x(q_{{\rm M},n},t_{{\rm c},n})$ measured in the simulations at successive crossing times, as illustrated by Fig.~\ref{fig:massprofile}. Importantly enough, with the right choice of this single free parameter $\beta$, theory matches the simulations rather well when the interior mass profile is at concern, in particular its close to a power-law behavior, $M_{\rm int}(x) \propto \sqrt{x}$. Such a power-law behavior cannot be obtained when the effect of the background is not taken into account in the dynamics. Indeed, in this case, the interior mass is underestimated as soon as $n \ga 5$ (this is not shown on Fig.~\ref{fig:massprofile}, for simplicity): as shown in previous section, when $\omega=0$ or in the toy model case, the projected density is expected to present a logarithmic singularity for small values of $x$.

Yet, it is interesting to see the differences in phase-space between various approximations when $3 \leq n \leq 6$: on bottom right panel of Fig.~\ref{fig:evolvpha}, the blue dotted curve assumes $\omega=0$, while the pink dashed one corresponds to the toy model given by equations (\ref{eq:xnoqmax}) and (\ref{eq:vnoqmax}). At third crossing time, the central shape of the ${\cal S}$ between the two extrema of the velocity is well described by the theory, whichever approach is considered. Obviously, this is not true for the remote tails, as expected. 

However, although the shape of the system at third crossing time is well described by the analytical calculations, the crossing time itself is slightly underestimated by the theory. This is illustrated by top panel of Fig.~\ref{fig:crossingtime}. In the case $\omega >0$, the disagreement between theory and measurements remains significant for $3 \leq n \la 6$, which corresponds to the time interval where our description of the dynamics is expected to be only marginally correct (bottom panel of Fig.~\ref{fig:crossingtime}). In the case $\omega=0$ or in the toy model, theory soon presents a plateau at late times that overestimate the true crossing time.  

We therefore have to set a time shift between the theory and the simulation in order to synchronize them. Equivalently, this simply consists in comparing theory to simulation at exact crossing times, as performed in Figs.~\ref{fig:evolvpha} and \ref{fig:evolvphaad}. This time shift is due to the fact that the effective density at the center of the system is slightly too large for the theoretical model, hence accelerating the rotation of the ${\cal S}$ in phase space.  This is not very surprising if one remembers the discussion after equation (\ref{eq:qv}) in \S~\ref{sec:omegaeq0}. Indeed, passing from the first crossing time to the second one can be performed accurately because the segment of time during which the system is multivalued is short, making our ballistic approximation rather accurate (when the system becomes monovalued again, the calculation of the force becomes exact again, as we get back to the Zel'dovich regime). When the system gains some spatial extension (limited by our cut-off at $q_{\rm M}$ corresponding to the extrema of the velocity), the relative amount of time between two crossings spent in the multivalued regime increases, which makes our calculation of the cumulated force exerted on a mass element less accurate. Yet, our estimate of the crossing times remains rather good as illustrated by top panel of Fig.~\ref{fig:crossingtime}. Note however that the effects of the warm nature of the initial waterbag on the value of the crossing time are felt quite early in the simulations: comparing the $v_{\rm max}=0.0003$ to the $v_{\rm max}=0.001$ simulation suggests that the simulations tend, at some point, to overestimate the true value of $t_{{\rm c},n}$. Comparison of theory and simulations at late times is therefore not fully meaningful. Strictly speaking, the numerical experiments can be trusted only up to $n=6$, where the two simulations still coincide with each other in terms of crossing times, or at best up to $n \sim 8$ for the $v_{\rm max}=0.0003$ simulation, in terms terms of interior mass and contamination at small scales by the presence of a warm core (Fig.~\ref{fig:massprofile}).

Even though the crossing time is underestimated by the theory for $3 \leq n \la 6$, the shape of the center ${\cal S}$ remains in agreement with the simulated one for up to 11 crossings, as illustrated by Fig.~\ref{fig:evolvphaad}, if one takes into account the presence of the background, $\omega > 0$. However, this does not mean that theory approaches the true solution beyond $n=8$ (strictly speaking, $n=6$, from the discussion just above), since the simulation is increasingly contaminated in the center by its warm nature. 

For $n \ga 12$, the agreement between the theory and the simulation worsens. At the last crossing time considered, $n=14$ (lower-left panel of Fig.~\ref{fig:evolvphaad}), the curvature of the tails of the red curve reverses, which is the signature of an instability that results in a change of sign of coefficient $a_n$, as already discussed in point (d) of \S~\ref{sec:recufor}.   If such an instability was present, the system would most likely build up a small flat core. To decide whether this instability is physical is not possible because, firstly, our post-Lagrangian approach is only approximate and, secondly, we do not have access to cold enough initial conditions in the simulations to prove that this instability would appear in the real system. 

To check even more accurately the validity of our iterative procedure, a powerful test consists in analysing the properties of the  {\em phase-space energy distribution function}, $f_E(E)$, which corresponds to the average of the phase-space density per energy level. Indeed, it has been shown in \cite{CT12} (hereafter, CT) that this function becomes nearly stationary for $n \ga 3$ when studied as a function of $E-E_{\rm min}$, where $E_{\rm min}$ is the (time dependent) minimum of energy coinciding with the minimum of the potential.  Furthermore, it was found by CT that the logarithmic slope of $f_E(E)$ was consistent, even at late times, with the one predicted at crossing times for small values of $E-E_{\rm min}$. 
\begin{figure*}
\centerline{\hbox{
\psfig{file=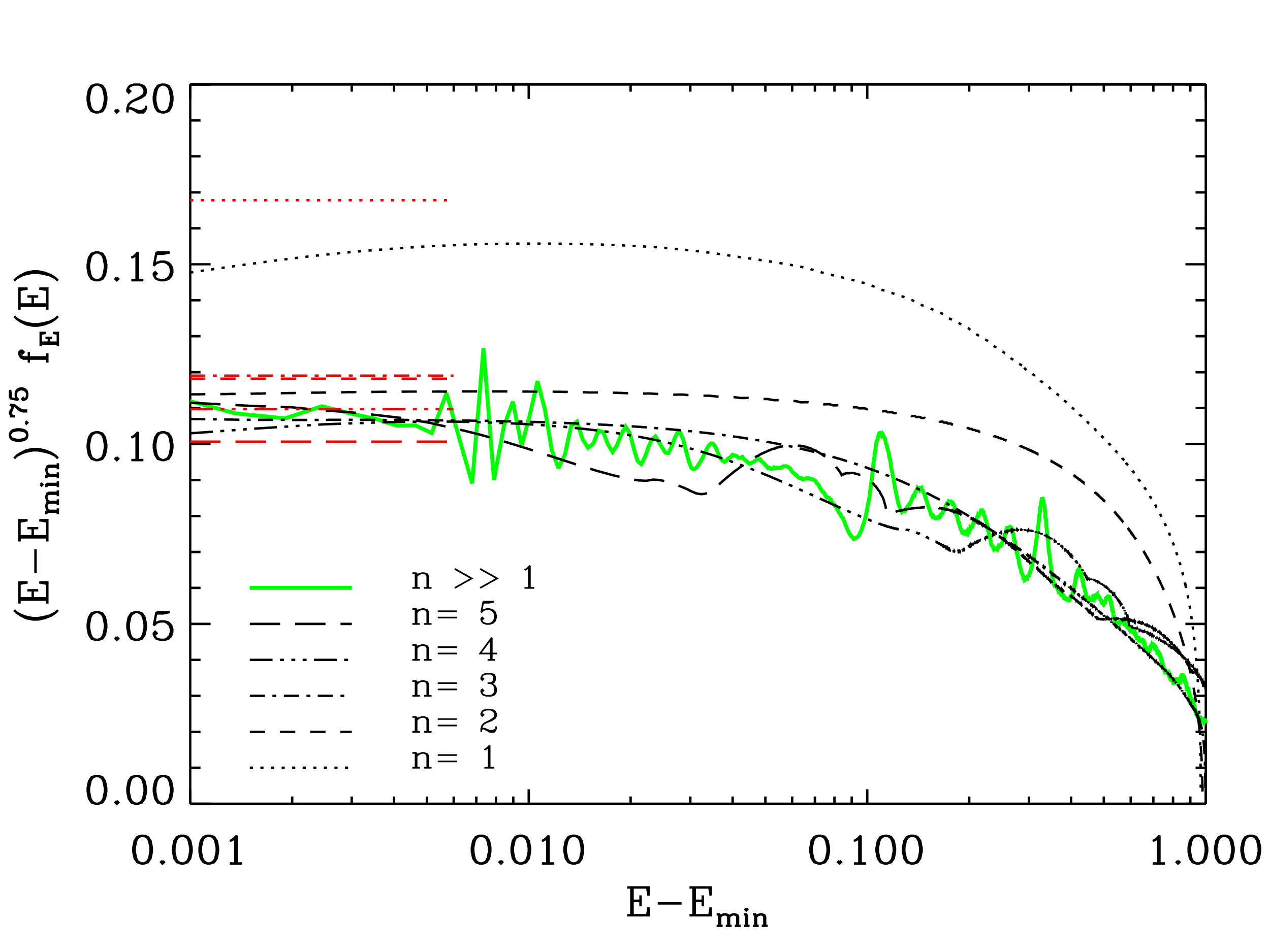,width=8.6cm}
\psfig{file=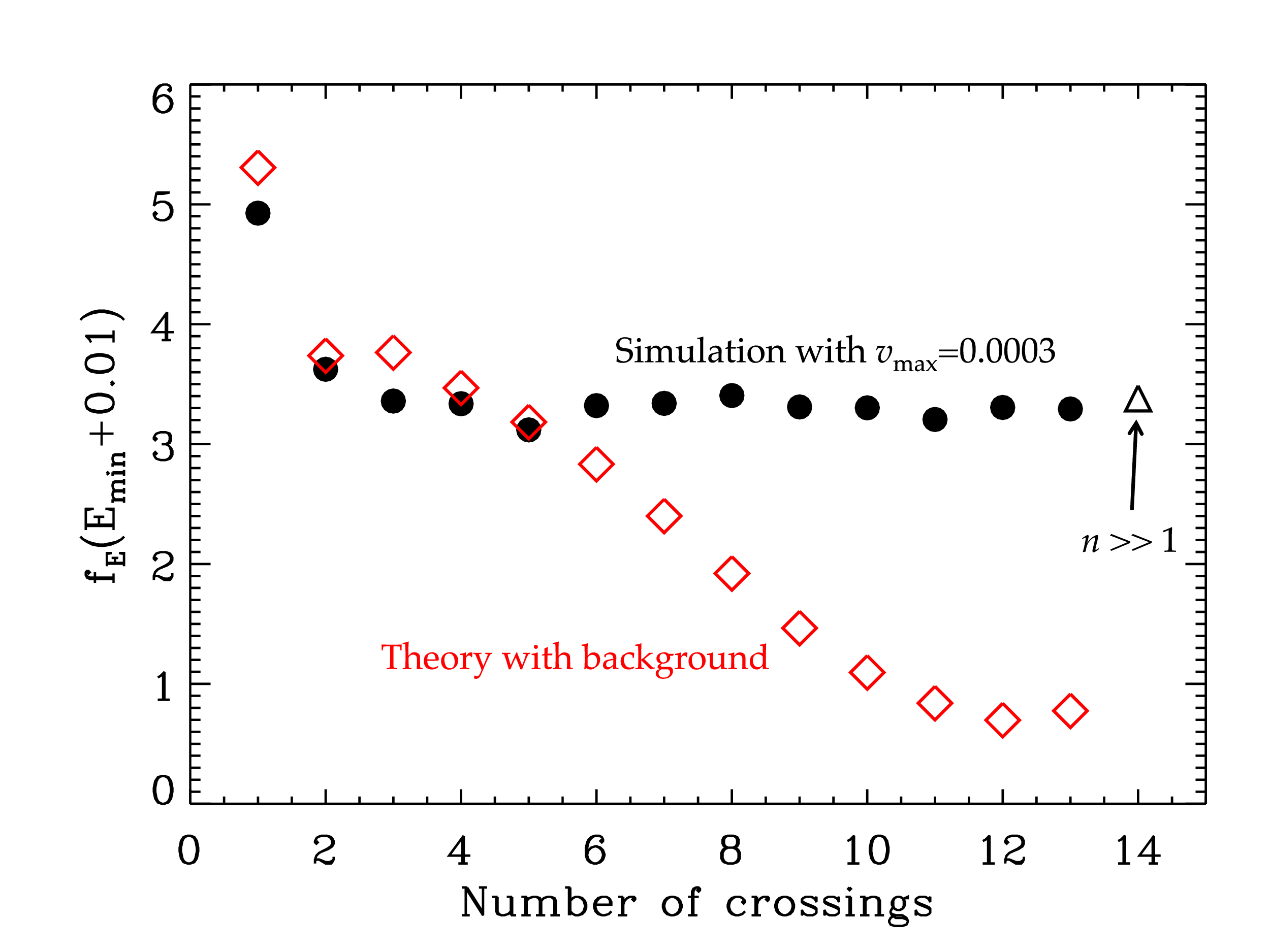,width=8.6cm}}}
\caption[]{Phase-space energy distribution function: comparison of theory to the simulation with $v_{\rm max}=0.0003$. The {\em left panel} shows the quantity $(E-E_{\rm min})^{3/4} f_E(E)$ as a function of specific energy at the first 5 crossing times as well as the most evolved time of the simulation, $t_{\rm s}=50$. The straight red lines with different patterns correspond to the theoretical prediction at each crossing time given by our iterative procedure with background taken into account. Each line has to be compared to the curve with the same pattern. The {\em right panel} focuses on a particular value of the energy, $E-E_{\rm min}=0.01$. The phase-space energy distribution function is shown as a function of number of crossings and is compared to the normalization given by the theory in the regime $E \ll 1$.}
\label{fig:fofefig}
\end{figure*}
Function $f_E(E)$ is defined as
\begin{equation}
f_E(E) \equiv \lim_{\delta E\rightarrow 0} \frac{\int_{E(x,v) \in [E,E+\delta E]} f(x,v)\ {\rm d}x\ {\rm d}v}{\int_{E(x,v) \in [E,E+\delta E]}\ {\rm d}x\ {\rm d}v},
\label{eq:fofedef}
\end{equation}
where 
\begin{equation}
E(x,v) \equiv \frac{1}{2} v^2+\phi(x)
\end{equation}
is the specific energy and $\phi(x)$ is the gravitational potential.  For systems where the phase-space density depends only on energy, one thus has $f(x,v)=f_E[E(x,v)]$.  At crossing times, equations (\ref{eq:recux}) and (\ref{eq:recuv}) imply a potential of the form
\begin{equation}
\phi(x) \simeq E_{\rm min}+\phi_0 x^{4/3}=E_{\rm min}+\phi_0 a_n q^4, \quad x,q \ll 1,
\label{eq:monphi}
\end{equation}
with
\begin{equation}
\phi_0=\frac{3}{2} a_n^{-1/3}.
\label{eq:monphi0}
\end{equation}
The specific energy can be easily computed as a function of $q$,
\begin{equation}
E  \simeq E_{\rm min}+\frac{1}{2} b_n^2 q^2, \quad q \ll 1.
\label{eq:enersmallq}
\end{equation}
 A convenient way to rewrite equation (\ref{eq:fofedef}) is
\begin{equation}
f_E(E)=2 \rho(q) \left[ \frac{\partial E}{\partial q}\right]^{-1} \left[ \frac{\partial J}{\partial E} \right]^{-1},
\label{eq:fofesi}
\end{equation}
where $J(E)$ is the action given by
\begin{equation}
J(E) \equiv \oint_{E(x,v)=E} v {\rm d}x=4 \int_0^{x_{\rm max}} \sqrt{2[E-\phi(x)]}{\rm d}x,
\end{equation}
and $x_{\rm max}$ is such that $E\equiv \phi(x_{\rm max})$. For a potential of the form $\phi-E_{\rm min} =\phi_0 x^{\beta_\phi}$, we have
\begin{equation}
J(E)=\frac{ 4 \sqrt{2\pi} \beta_\phi \Gamma(1+\frac{1}{\beta_\phi}) \phi_0^{-\frac{1}{\beta_\phi}} (E-E_{\rm min})^{\frac{1}{2}+\frac{1}{\beta_\phi}}}{(2+\beta_\phi) \Gamma(\frac{1}{2}+\frac{1}{\beta_\phi})}
\label{eq:jofe}
\end{equation}
(see, e.g., CT). From equations (\ref{eq:monphi}), (\ref{eq:monphi0}), (\ref{eq:enersmallq}), (\ref{eq:fofesi}) and  (\ref{eq:jofe}), we obtain
\begin{eqnarray}
f_E(E) & \simeq & \frac{(3/2)^{3/4} \Gamma(5/4)}{2 \sqrt{\pi} \Gamma(7/4)} \frac{(E-E_{\rm min})^{-3/4}}{a_n^{1/4} b_n}, \nonumber \\
& & \quad \quad \quad \quad \quad \quad \quad \quad  \quad \quad \quad E-E_{\rm min} \ll 1. \label{eq:powe}
\label{eq:fofefinal}
\end{eqnarray}
In particular, we should have, for $n \ga 2$, 
\begin{equation}
a_n^{1/4} b_n \simeq {\rm constant},
\label{eq:propanbn}
\end{equation}
according to the measurements of CT. 

Figure~\ref{fig:fofefig} shows $(E-E_{\rm min})^{3/4} f_E(E)$ as a function of specific energy (left panel) as well as $f_E(E_{\rm min}+0.01)$ as a function of time (right panel). The method employed to compute $f_E(E)$ from equation (\ref{eq:fofedef}) is explained in details in \cite{CT12}. 2500 linear bins were used in the interval $[E_{\rm min},E_{\rm max}]$, where $E_{\rm min}$ and $E_{\rm max}$ are respectively the minimum and the maximum of energy inside the waterbag. 
Left panel of Fig.~\ref{fig:fofefig} confirms the results of CT, namely the power-law behavior $f_E(E) \propto (E-E_{\rm min})^{0.75}$. However, the normalization of function $f_E(E)$, in particular equation (\ref{eq:propanbn}), is reproduced by our analytic calculations only up to $n \simeq 6$ (right panel). After this, the energetic properties of the approximate system deviate significantly from the simulation and probably from the true solution. Indeed, the close to stationary behavior of function $f_E(E)$ at late times --modulo the small scales fluctuations due to the cold nature of the system-- seems to be a robust numerical result according to the convergence analysis performed in \citet{CT12}. In particular, the deviation between theory and measurements observed on right panel of Fig.~\ref{fig:fofefig} cannot be attributed to the fact that the simulation is not exactly cold. 

\section{Summary and discussion}
\label{sec:conclusion}
In this work, we have studied analytically the evolution of an initially smooth and cold system following Vlasov-Poisson dynamics in one dimension. We focused on the central part of the system, which presents an ${\cal S}$ shape in phase-space. We approximated the equations of motion of the ${\cal S}$ using a Lagrangian perturbative approach, which allowed us to describe its dynamics beyond crossing. Then, by iterating the procedure,  we followed the system during several crossing times. The ${\cal S}$ rotates in phase-space and its central part contracts under the action of self-gravity. Its tails feed a halo which contributes to the dynamics of the ${\cal S}$ as an increasing harmonic background. With this simple recipe, in which the phase-space coordinates of the curve representing the ${\cal S}$ are just third order polynomials of the Lagrangian coordinate $q$, we were able to describe rather accurately the evolution of the central part of the system during several orbits and the establishment of a close to power-law behavior for the projected density, as measured in numerical simulations.

From these conceptually simple but yet cumbersome calculations, it seems that the detailed dynamics of the shape of the ${\cal S}$ completely determines the very existence and the nature of the singularity at the center of the system, if one assumes that the tails of the ${\cal S}$ feed a nearly stationary halo. This result, which can be only derived by following the exact phase-space distribution function of the center of the system at the microscopic level, is clearly incompatible with any coarse grained approach, such as methods using entropy maximization, unless they account for the non trivial constraints induced by the dynamics of this ${\cal S}$. 

In the continuous and cold limit considered here, coarse graining can indeed be used to define a macro-state that consists of a region of phase-space large enough compared to the typical separation between two close elements of the spiral composing the system, but small enough compared to the extension of the system. Entropy maximization assumes that small regions of phase-space corresponding to ``micro-states'', initially very far from each other, hence uncorrelated statistically, can end up in the same macro-state as consequence of mixing. Entropy is a count of the number of combinations of micro-states leading to the same macro-states. Maximizing the entropy is thus equivalent to finding the most likely state of the system given the lack of knowledge we have about it. Additional constraints, which on the contrary reflect our knowledge about the system, are used to set limits to the statistical freedom expressed here in terms of number of microstates. 

In Lynden-Bell theory \citep[][]{LyndenBell}, conservation of the total mass, total energy and of local phase-space volume represent such constraints. However, Lynden-Bell theory has been proven to have a limited predicting power.  For instance, although it is partly successful \citep[see, e.g.,][]{Yamaguchi2008}, it fails to reproduce in detail the steady state of many one-dimensional systems \citep[see, e.g.,][]{Joyce2011}. Additional insights on the dynamical properties of the system can however significantly improve it, but they always relie on a particular treatment at the center of system, for instance through a tunable parameter defining the gravitational potential depth \citep[see, e.g.][]{Williams2010,Carron2013} or a second component of the phase-space distribution function \citep[see, e.g.,][]{Pontzen2013}. 

Our calculations suggest, in the same line of thought, that it is needed in the cold and one dimensional case studied in this article to add fully deterministic constraints on the phase-space density in the vicinity of the center of the system to have any hope for an entropic prescription to produce the correct quasi-stationary solution. Indeed, in the very center of the system, there is no mixing, at least during a significant amount of time after which the system has already reached a metastable state. Furthermore, the spiral structure that builds up during the course of the dynamics is well defined \citep[see, e.g.,][]{Gurevich1995,Alard2013}. In particular, a given small patch of phase-space does not contain random parts of such a spiral: this may actually render the concept of entropy maximization questionable in full when considering the system examined in this article. 

Indeed, it is tempting to believe that there should be a random component at play during the mixing process for the maximum entropy method to really make sense. When the system is composed of particles, Poisson noise combined with local encounters plays the role of such a random component. Hence, the long term evolution of the system leads in this case to the thermodynamic solution that can be obtained by maximizing the entropy \citep[see, e.g.,][and references therein]{Lyndenbell1968,Rybicki1971,Chavanis2006}. In the continuous limit, these is no such a random component, unless the system becomes chaotic or there are seeds in the initial distribution that stand for small units randomly distributed. Interestingly, in the cold dark matter paradigm, the formation of a dark matter halo is not a monolitic process but results from successive mergers of smaller structures formed earlier. In this respect, one might imagine that the randomness of such a process could be approached efficiently with a maximum entropy method. 

As a final note on entropy maximization, is also important to point out another problem in the continuous limit. The definition of micro-states that would be ``statistically equivalent to each other'' is not unique, because there is  {\em a priori} no obvious elementary unit to deal with, at variance with systems of particles. Moreover, as already mentioned in the introduction, there are many ways to define coarse graining that lead different answers for the maximum entropy solution. This means that there is no unique definition of entropy, which complicates even furthermore the problem \citep[see, e.g.,][and references therein]{Tremaine1986,Chavanis2006}. Decomposing the phase-space density into a set of indistinguishable particles interacting with each other gravitationally and maximizing the entropy of such a system\footnote{In this latter case, one can directly define from first principles a statistical entropy, equivalent to the one used by physicists, without necessarily resorting to coarse-graining \citep[see, e.g.,][]{Jaynes1957}.} leads, as said above, to the thermodynamic solution which is not, in general, the physical solution of interest in the continuous limit, for which it is known that there exist infinitely many stationary and stable solutions to which the system can converge \citep[see, e.g.,][and references therein]{Campa2009}. 

The post-Lagrangian perturbative method developed in this work can certainly be applied to spherical systems to give clues about the establishment of so-called universal profiles of dark matter halos. In this case, phase-space is three-dimensional: the radial position, the radial velocity and the angular momentum. Because the angular momentum is an invariant of the motion, the problem presents at the end formally the same level of complexity as the two-dimensional phase-space case we studied in this paper. Therefore, the calculations should not be significantly more difficult to perform. However, the interest of repeating the exercise in such a restrictive geometry is limited since it is known that pure spherical dynamics is unrealistic \citep[see, e.g.,][]{Huss1999}.

It might however be possible to apply, at least partly, the method to ellipsoidal systems, although the task would become complex despite the high level of symmetry of these configurations. Indeed, in this case, there are up to three directions in which the system can collapse successively, depending on the nature of initial conditions. While one dimensional approximations of the dynamics as developed in the present work would certainly be relevant during the early stages of the evolution of the system, it might be difficult to account for the expected non trivial couplings occurring between the different directions of motion. 

The extension of the method to the general three dimensional case is of course even trickier due to the complexity of the structure of the singularities that build up during the course of dynamics and can be classified according to catastrophe theory \cite[see, e.g.][and references therein]{Arnold1982, Arnold1984,Shandarin1989,Hidding2014}. Indeed, in one dimension, there are only two types of singularities: the first kind corresponds to the case for which the central part of the ${\cal S}$ shape is vertical in phase-space, leading to equation (\ref{eq:rhosing}), i.e. a behavior with $\rho(x) \propto x^{-2/3}$. The second kind of singularity corresponds to the case where the curve supporting the phase-space density is also locally vertical but with non zero curvature. In this latter case, the singularity is of the form $\rho(x) \propto x^{-1/2}$. While the first kind of singularity only appears at crossing times in the center of the system, the second kind continues existing after its genesis. Creation of singularities of the second kind results from the rotation of the ${\cal S}$ in phase-space, which transforms immediately a singularity of the first kind into 2 singularities of the second kind standing for the handles of the ${\cal S}$ shape. 

In the present work, we described the system in a perturbative way in the region where the singularity of the first kind appears. The singularities of the second kind do not really matter from the dynamical point of view except when they correspond to the handles of the ${\cal S}$. In higher number of dimensions, the number of species of singularities increases and their structure is more complex \cite[see, e.g.][]{Hidding2014}. Isolating the singularities which matter from the dynamical point of view and implementing a realistic description of the dynamics in their vicinity with our post Lagrangian approach seems a challenging task. Such a task would become practically impossible if one would have to go beyond third order in the perturbative description. Solving first the ellipsoid case might help quite a lot, because the dynamical setting can to a large extent be locally reduced to the ellipsoid case by working in the coordinate frame that diagonalizes the deformation tensor. This is indeed what Zeldovich approximation suggests us \citep[][]{Zeldo}, at least during the early stages of the dynamics. However, the resolution of the problem does not reduce to the formation and evolution of single objects around initial singularities. Indeed, one then has to take into account mergers between various structures that create complex composite systems. The proper dynamical description of such merger events at the level of accuracy intended here is highly non trivial and goes well beyond the simplified dynamical model discussed in this article. 
\section*{Acknowledgements}
I would like to thank Jihad Touma, who originated this project about five years ago by suggesting to study the evolution of the central ${\cal S}$ in a way analogous to the method developed in this work, but by using a simplified numerical toy model. I also discussed similar ideas more recently with James Binney during Gravasco trimester at Institut Henri Poincar\'e. I would also like to thank Christophe Alard, J\'er\^ome Perez and Scott Tremaine for useful discussions as well as Pierre-Henri Chavanis for enlightenments on the maximum entropy methods. Most of the calculations presented in this article have been performed with the help of {\tt Mathematica} and I am most grateful to Guillaume Faye for training me with this software. This work has been funded in part by ANR grant ANR-13-MONO-0003.

\appendix
\section{Calculation of the equations of motion: details}
\label{app:cal}
\subsection{The expression for the different event times}
\label{app:timesdetails}
In this appendix, we provide detailed analytical expressions for the various event times as well as their second order series expansion, which is used for the calculation of the correction to the harmonic (or ballistic if $\omega=0$) approximation of the trajectory, $[x_{\rm b}(q,h),v_{\rm b}(q,h)]$. 

The positive Lagrangian coordinate corresponding to the leftmost position of the handle of the ${\cal S}$ shape is defined by equation (\ref{eq:defqc}). It reads
\begin{equation}
q_{\rm c}(h)=\sqrt{\frac{b \sin( \omega h)}{3 [a \omega \cos( \omega h)+ c \sin( \omega h)]}}.
\label{eq:qcexpr}
\end{equation}
This equation can be inverted into the event time
\begin{equation}
h_{\rm c}(q)=\frac{1}{w} \arccos \left[ \frac{b-3 c q^2}{\sqrt{(b - 3 c q^2)^2 + 9 a^2 \omega^2 q^4}}\right],
\label{eq:hcexp}
\end{equation}
or, in the absence of background, 
\begin{equation}
h_{\rm c}(q)=\frac{3 a q^2}{b-3 c q^2}, \quad w \rightarrow 0.
\end{equation}
Remind that this time marks the transition when, due to clockwise rotation of the curve $[x_{\rm b}(q,t),v_{\rm b}(q,t)]$, a point being of  ``$q_1$'' kind becomes of ``$q_0$'' kind on Fig.~\ref{fig:explain}.

Inverting equation ${\hat q}_{\rm c}(h)=q$, with ${\hat q}_{\rm c} \equiv 2 q_{\rm c}$ (equation \ref{eq:defhatqc}) allows us to define another critical event time
${\hat h}_{\rm c}(q) \leq h_{\rm c}(q)$,
\begin{equation}
{\hat h}_{\rm c}(q)=\frac{1}{\omega}\arccos \left[\frac{4b-3 c q^2}{\sqrt{(4b - 3 c q^2)^2 + 9 a^2 \omega^2 q^4}}\right],
\label{eq:hathcexp}
\end{equation}
or, in the absence of background,
\begin{equation}
{\hat h}_{\rm c}(q)=\frac{3 a q^2}{4 b- 3 c q^2}, \quad w \rightarrow 0.
\end{equation}
This time marks the transition when a point becomes of ``$q_2$'' kind, that is when one passes from single valued solution to triple valued one, or, in other words, when the point $x(q,t)$ crosses for the first time another trajectory $x(q',t)$.  

Depending on the value of $h$ considered, both $q_{\rm c}$ and ${\hat q}_{\rm c}$ can have larger magnitude than $q_{\rm M}$. Indeed, even though $|q| \leq {\hat q}_{\rm c}$, the solutions $q'$ of the equation $x_{\rm b}(q',h)=x_{\rm b}(q,h)$ can be outside the system,  $|q'| > q_{\rm M}$.  This defines two new important time events for a Lagrangian point $q$, which correspond respectively to the two solutions (in time), $h_{-}(q) \leq h_{+}(q)$,  of the equation $x(q,h)=x(q_{\rm M},h)$:
\begin{eqnarray}
h_{\pm}(q) &=& \frac{1}{\omega} \arccos\left[ \frac{b-c \Delta_{\pm}}{\sqrt{a^2 \omega^2{\Delta_{\pm}}^2+(b-c \Delta_{\pm})^2}}\right],\\
\Delta_{\pm} &\equiv &q^2+q_{\rm M}^2\pm q_{\rm M} |q|,
\label{eq:hpm}
\end{eqnarray}
or, in the absence of background, 
\begin{equation}
h_{\pm}(q)=\frac{a \Delta_{\pm}}{b-c \Delta_{\pm}}.
\end{equation}
The time $h_{-}(q)$ defines the instant when the number of valid solutions of the equation $x(q',h)=x_0$ with $x_0=x(q,h)$ passes from 3 (including $q$ itself) to 2 (one solution is outside the system). The time $h_{+}(q)$ defines the instant when the number of valid solutions of the equation $x(q',h)=x_0$ with $x_0=x(q,h)$ passes from 2  (including $q$ itself) to 1 (two solutions are outside the system). 

To understand better what happens during the trajectory of a matter element, Figure~\ref{fig:charactimes} shows the various event times, ${\hat h}_{\rm c}$, $h_{\rm c}$ and $h_{\pm}$ as functions of $q$, as well as their second order series expansion in $q$ that we shall use in the subsequent calculations:
\begin{eqnarray}
{\hat h}_{\rm c}(q) &=&\frac{3 a q^2}{4b}+{\cal O}(q^4), \label{eq:hathcap} \\
h_{\rm c}(q) & = & \frac{3 a q^2}{b}+{\cal O}(q^4), \label{eq:hcap} \\
h_{\pm}(q) &=& \frac{1}{\omega} \arccos\left( \frac{b-c q_{\rm M}^2}{\sqrt{T}} \right) \pm \frac{ a b q_{\rm M}}{T}|q|+ \nonumber \\
& & \frac{a b^2 (b - c q_{\rm M}^2)}{T^2}|q|^2+{\cal O}(q^3), \label{eq:hpmexp}
\end{eqnarray}
with 
\begin{equation}
T=(b-c q_{\rm M}^2)^2+a^2 q_{\rm M}^4 \omega^2.
\label{eq:eqforT}
\end{equation}
For completeness the no background limit of equation (\ref{eq:hpmexp}) reads
\begin{eqnarray}
h_{\pm}(q) &=&\frac{a q_{\rm M}^2}{b-c q_{\rm M}^2}\pm \frac{ a b  q_{\rm M}}{(b- c q_{\rm M}^2)^2} |q|+ \nonumber \\ 
& & \frac{a b^2}{(b- c q_{\rm M}^2)^3} |q|^2 +{\cal O}(q^3), \quad \omega \rightarrow 0. \label{eq:hpmexpw0}
\end{eqnarray}
\subsection{The expression for the acceleration in different phases of the motion}
\label{sec:forcecalc}
We now compute the force exerted on a particle of mass unity belonging to the system in each of the time intervals delimited by the critical events described in previous section. These latter are ranked in the following order, ${\hat h}_{\rm c} \leq h_{\rm c} \leq h_- \leq h_+$, if $q$ is small enough. 

When $q \geq {\hat q}_{\rm c}$, the force is monovalued and given by equation (\ref{eq:accvelt2}), while it is given by equation (\ref{eq:massintformula}) when the system becomes multivalued, that is when $q < {\hat q}_{\rm c}$. The parameters $q_0$, $q_1$ and $q_2$, ranked in increasing magnitude order such that $|q_0| \leq q_{\rm c} \leq |q_1| \leq |q_2| \leq {\hat q}_{\rm c}$, are solutions of the equation
\begin{equation}
q^3-\frac{3}{4} q {\hat q}_{\rm c}^2={q'}^3-\frac{3}{4} {q'} {{\hat q}_{\rm c}}^2,
\label{eq:xqxqp}
\end{equation}
or
\begin{equation}
(q-q')[q^2+qq' +{q'}^2-3 {\hat q}_{\rm c}^2/4]=0.
\label{eq:soluq}
\end{equation}
The non trivial solutions of this equation are 
\begin{equation}
q'=-\frac{q}{2}\pm\frac{1}{2}\sqrt{3({\hat q}_{\rm c}^2-q^2)}.
\end{equation}
In particular, on Fig.~\ref{fig:explain}, we have the circular relations,
\begin{eqnarray}
q_1 & = &-\frac{q_0}{2}-\frac{1}{2}\sqrt{3({\hat q}_{\rm c}^2-{q_0}^2)}, \label{eq:q1}\\
q_2 & = & -\frac{q_0}{2}+\frac{1}{2}\sqrt{3({\hat q}_{\rm c}^2-{q_0}^2)}, \label{eq:q2}\\
q_0 & = & -\frac{q_1}{2}-\frac{1}{2}\sqrt{3({\hat q}_{\rm c}^2-{q_1}^2)}, \label{eq:q3}\\
q_2 & = & -\frac{q_1}{2}+\frac{1}{2}\sqrt{3({\hat q}_{\rm c}^2-{q_1}^2)}, \label{eq:q4}\\
q_0 & = & -\frac{q_2}{2}+\frac{1}{2}\sqrt{3({\hat q}_{\rm c}^2-{q_2}^2)}, \label{eq:q5}\\
q_1 & = & -\frac{q_2}{2}-\frac{1}{2}\sqrt{3({\hat q}_{\rm c}^2-{q_2}^2)}. \label{eq:q6}
\end{eqnarray}
Now, we are ready to compute the force exerted on a particle of mass unity at any instant of the trajectory $[x_{\rm b}(q,h),v_{\rm b}(q,h)]$. First we consider the case $q \leq q_{\rm M}/2$, where $q_{\rm M}/2$ marks the transition $h_{\rm c}(q)=h_-(q)$. The force presents a different behavior for each phase delimited by the event times plotted on Fig.~\ref{fig:charactimes}, as illustrated by top panel of Fig.~\ref{fig:forcetime} and discussed in \S~\ref{sec:phases}:
\begin{enumerate}
\item{\em  Prior to first-crossing}, $h \leq {\hat h}_{\rm c}(q)$: the force is simply given by
\begin{equation}
F(q,h)=F_1(q,h)\equiv-2(q-q^3),
\end{equation}
exactly as in \S~\ref{sec:zeldo}.
\item{\em Between first-crossing and interior phase}, ${\hat h}(q)
  \leq h \leq h_{\rm c}(q)$: using equations (\ref{eq:massintformula}), (\ref{eq:mqsimp}) and e.g. (\ref{eq:q5}) and (\ref{eq:q6}), we obtain
\begin{eqnarray}
F(q,h) & =& F_2(q,h),  \\
F_2(q,h) &\equiv &-2(q-q^3)+ \nonumber \\
 & & 2\, {\rm sgn}(q) \left[1-\frac{3}{4} {\hat q}_{\rm c}^2(h)\right] \sqrt{3{\hat q}_{\rm c}^2(h)-3q^2}.
\end{eqnarray}
\item{\em Interior phase}, $h_{\rm c}(q) \leq h \leq h_-(q)$: the force reads
\begin{equation} 
F(q,h)=F_3(q,h)\equiv 2 q (2+q^2)-\frac{9}{2}{\hat q}_{\rm c}^2(h) q.
\label{eq:F3}
\end{equation}
Note for completeness that, in the limit $\omega \rightarrow 0$,
\begin{equation}
F_3(q,h)= 2 q (2 + q^2)-\frac{ 6 b t}{a+c t} q,\quad \omega \rightarrow 0. 
\label{eq:f3approx}
\end{equation} 
\item{\em interior phase, but a tail not contributing anymore}, $h_-(q) \leq h \leq h_+(q)$: the tails of the ${\cal S}$ contribute only partly to the force, which complicates its expression:
\begin{eqnarray}
F(q,h) & = & F_4(q,h), \\
F_4(q,h) & \equiv & 2 (q_{\rm M}-q_{\rm M}^3) {\rm sgn}(q)+\nonumber \\
& & \left[ 1-\frac{3}{4} {\hat q}_{\rm c}^2(h)\right] \left[ 3 q -
  {\rm sgn}(q) \sqrt{3{\hat q}_{\rm c}^2(h)-3q^2}\right]. \nonumber \\
& &
\end{eqnarray}
\item{\em back to the monovariate regime}, $h_+(q) \leq h$: the ${\cal S}$ shape has rotated in phase-space in such a way that its tails do not influence anymore the motion of its central part. The force is again equal to
\begin{equation}
F(q,h)=F_5(q,h)\equiv 2(q-q^3),
\end{equation} 
and we are ready to proceed until next crossing.
\end{enumerate}
When $|q| \geq q_{\rm M}/2$ (second panel of Fig.~\ref{fig:forcetime}), the phase (iii) disappears. A regime $h_-(q) \leq h \leq h_{\rm c}(q)$ now appears, where the force is
\begin{eqnarray}
F(q,h) &= & {\tilde F}_3(q,h), \\
{\tilde F}_3(q,h) &\equiv & 2 (q_{\rm M}-q_{\rm M}^3) {\rm sgn}(q)-\nonumber \\
& & \left[ 1-\frac{3}{4} {\hat q}_{\rm c}^2(h)\right] \left[ 3 q -
  {\rm sgn}(q) \sqrt{3{\hat q}_{\rm c}^2(h)-3q^2}\right]. \nonumber \\
& &
\end{eqnarray}

Since our calculations aim to be only accurate at third order in $q$, we can simplify the expressions of the force obtained at the various
times above. For instance, the correction to the velocity can written, for $q \leq q_{\rm M}/2$ and $h \geq h_+(q)$, 
\begin{eqnarray}
\frac{\partial g}{\partial h} & =& \int_0^{{\hat h}_{\rm c}(q)} F_1(q,h) {\rm d}h+\int_{{\hat h}_{\rm c}(q)}^{h_{\rm c}(q)} F_2(q,h) {\rm d}h+\nonumber \\
& &\int_{h_{\rm c}(q)}^{h_{-}(q)} F_3(q,h) {\rm d}h +\int_{h_{-}(q)}^{h_{+}(q)} F_4(q,h) {\rm d}h+\nonumber \\
& & \int_{h_+(q)}^{h} F_5(q,h) {\rm d}h.
\label{eq:myvdiff}
\end{eqnarray}
Because both ${\hat h}_{\rm c}(q)$ and $h_{\rm c}(q)$ are of the order of ${\cal O}(q^2)$, it is required to be only accurate at first order in $q$ and $h$ for $F_1$ and $F_2$, 
\begin{eqnarray}
F_1(q,h) & \simeq & -2 q, \\
F_2(q,h) & \simeq & -2 q + 2\, {\rm sgn}(q) \sqrt{\frac{4 b h}{a}-3 q^2}. 
\end{eqnarray}
On the other hand $F_3$ can not be Taylor expanded because $h_-(q) \sim {\cal O}(1)$. 
From equation (\ref{eq:hpmexp}) (or equation~\ref{eq:hpmexpw0}), one finds $h_+(q)-h_-(q) \sim {\cal O}(q)$, which means that we need second order in $q$ and in $h-h_-(q)$ for $F_4$. Setting
\begin{equation}
\theta=h-h_-(0),
\end{equation}
and performing the expansion at second order in $\theta$ for ${{\hat q}_{\rm c}}^2$:
\begin{equation}
\frac{3}{4}{{\hat q}_{\rm c}}^2\simeq q_{\rm M}^3+\frac{T}{ab} \theta + \frac{T [q_{\rm M}^2(c^2+a^2 \omega^2)-b c]}{a^2 b^2} \theta^2+{\cal O}(\theta^3),
\end{equation}
where $T$ is given by equation (\ref{eq:eqforT}), the series expansion of $F_4$ can be performed easily
\begin{eqnarray}
F_4(q,h) & \simeq & F_{4,1} q+F_{4,2}\, {\rm sgn}(q)\, [h-h_-(0)]+ \nonumber \\
& & F_{4,3}\, q\, [h-h_-(0)]+ F_{4,4}\, {\rm sgn}(q)\, q^2 + \nonumber \\
& & F_{4,5} \,{\rm sgn}(q)\, [h-h_-(0)]^2,
\label{eq:F4expan}
\end{eqnarray}
with
\begin{eqnarray}
F_{4,1} & = & 3(1-q_{\rm M}^2) , \\
F_{4,2} &= &\frac{(3 q_{\rm M}^2 -1) T}{a b q_{\rm M}},  \\
F_{4,3} & = & -\frac{3 T}{a b} , \\
F_{4,4} & = & \frac{3 (1-q_{\rm M}^2)}{4 q_{\rm M}},  \\
F_{4,5} &=& \frac{ T [(c^2+a^2 w^2) q_{\rm M}^2- b c] (3 q_{\rm M}^2-1)}{ a^2 b^2 q_{\rm M}}+ \nonumber \\
      & &  \quad \quad \quad \frac{T^2 (1+3 q_{\rm M}^2)}{4 a^2 b^2 q_{\rm M}^3}.
\end{eqnarray}
In equation (\ref{eq:F4expan}), we have kept terms up to $q^2$, $q \theta$ and $\theta^2$. 

Finally, obviously, $F_5$ cannot be simplified. 
\subsection{Calculation of the corrected motion}
\label{app:cormot}
In equation (\ref{eq:velfinalbig}), one has to estimate, for each of the phases of the motion, the primitives with time, $G_i(q,h)$, of the functions $F_i(q,h)$ calculated in previous appendix.  After some algebra, one obtains, at third order in $q$:
\begin{eqnarray}
G_1(q,h) &=&-2q h, \\
G_2(q,h) &=&-2 q h +\frac{a}{3b}{\rm sgn}(q) \left( \frac{4b}{a} h-3 q^2 \right)^{3/2} , \\
G_3(q,h) &=& 2 q \left(2+q^2-\frac{3 b c}{c^2 + a^2 \omega^2} \right) h+ \nonumber \\
              &   & \frac{6 a b \ln[ a \omega \cos(\omega h) + c \sin(\omega h)]}{c^2 + a^2 \omega^2} q, \\
              &= & 2 q \left( 2 -\frac{3b}{c}+ q^2 \right) h  + \frac{ 6 a b \ln(a+c h)}{c^2} q \nonumber \\
              &   & \quad \quad \quad \hskip 2cm \omega \rightarrow 0 \\
G_4(q,h) &=& F_{4,1}\,q\, [h-h_-(0)] +\nonumber \\
              &   &\frac{1}{2} F_{4,2}\,{\rm sgn}(q)\,[h-h_-(0)]^2+\nonumber \\
              &   & \frac{1}{2}F_{4,3}\,q\,[h-h_-(0)]^2 +\nonumber \\
              &   & F_{4,4}\,{\rm sgn}(q)\,q^2\,[h-h_-(0)]+\nonumber\\
              &   & \frac{1}{3}F_{4,5}\,{\rm sgn}(q)\,[h-h_-(0)]^3,\\
G_4(q,h) & =& 2 (q -q^3) h.
\end{eqnarray}
Note that, even though the analytical result provided by equation (\ref{eq:velfinalbig}) --from which a few terms remain to be dropped out in order to stay at third order in $q$-- was computed in the regime $q \leq q_{\rm M}/2$, it {\em stands as well for} $q \geq q_{\rm M}/2$, by continuity. Indeed, if we would perform all the calculations in this regime, by taking into account the modifications in phases (ii) and (iii) and the force ${\tilde F}_3$ instead of $F_3$ as discussed in \S~\ref{sec:forcecalc}, at the end, we would obtain exactly the same expressions at third order in $q$. This is not surprising because there is no reason of having a discontinuity at $q = q_{\rm M}/2$.

Turning to the position, we can happily drop the terms $i=1$ and $i=2$ in the sums in second and third line of equation (\ref{eq:myxfinalbig}), since they are beyond third order. We still need functions $H_3$, $H_4$ and $H_5$ of which we would like to keep the part that contributes up to third order in $q$. There is no particular difficulty for $H_4$ and $H_5$: 
\begin{eqnarray}
H_4(q,h) &=& \frac{1}{2}F_{4,1}\,q\, [h-h_-(0)]^2 +\nonumber \\
              &   &\frac{1}{6} F_{4,2}\,{\rm sgn}(q)\,[h-h_-(0)]^3,\\
 H_5(q,h) & =& (q -q^3) h^2,
\end{eqnarray}
where we used the fact that in the range $[h_-(q),h_+(q)]$, $h-h_-(0) \sim {\cal O}(q)$.

Unfortunately, analytical calculation of function $H_3$ is beyond our skill, except in the limit $\omega \rightarrow 0$:
\begin{eqnarray}
     H_3(q,h)&=& -\frac{6 a b}{c^2} q h  + \left(2-\frac{3 b q}{c}\right) q h^2 +\nonumber \\
              &  & \frac{6 a b}{c^3} q (a + c h) \ln(a + c h) + q^3 h^2,  \nonumber \\
               & & \quad \quad \quad \hskip 2cm \omega \rightarrow 0.
\end{eqnarray}
To consider the case $\omega \neq 0$, we first notice that $H_3(q,h)$ needs to be evaluated at $h_3=h_{\rm c}(q)$ and $h_4=h_-(q)$. The smooth nature of function $G_3(q,h)$ in the domain $[0,h_-(q)]$ allows us to define function $H_3(q,h)$ as
\begin{equation}
H_3(q,h)\equiv \int_0^h G_3(q,h') {\rm d}h'.
\end{equation}
For evaluating $H_3(q,h)$ at $h_3=h_{\rm c}(q)$, one just need to Taylor expand $G_3$ at leading order in $h$ and integrate it to obtain
\begin{equation}
H_3(q,h) \simeq \frac{6 a b \ln( a \omega )}{c^2 + a^2 \omega^2}q h, \quad h \ll 1.
\end{equation}
The estimate of $H_3(q,h)$ at $h_4=h_-(q)$ can be performed as follows
\begin{eqnarray}
H_3(q,h) & = & I_1+I_2, \label{eq:H3deal}\\
I_1 &\equiv &\int_0^{h_-(0)} G_3(q,h) {\rm d}h \\
     & = &  Y+h_-(0)^2 q \left( 2 + q^2 - \frac{3 b c}{c^2+a^2 \omega^2} \right) \\
I_2 & \equiv & \int_{h_-(0)}^{h} G_3(q,h) {\rm d}h \\
     &\simeq &  \frac{3 q[b c - q_{\rm M}^2 (c^2+\omega^2)]}{c^2 + a^2 \omega^2} [h-h_-(0)]^2 + \nonumber \\
     & & q \left(2- \frac{3b c}{c^2 + a^2 \omega^2} \right) [ h^2 - h_-(0)^2]+\nonumber \\
     & & \frac{6 a b q}{c^2+a^2 \omega^2} \ln\left( \frac{a b \omega}{\sqrt{T}}\right)[h-h_-(0)],
\end{eqnarray}
with
\begin{equation}
Y \equiv \int_0^{h_-(0)} \frac{6 a b \ln[a \omega (\cos \omega h) + c \sin(\omega h)]}{c^2 +a^2 \omega^2} {\rm d}h.
\label{eq:eqforY}
\end{equation}
The quantity $Y$ has to be estimated numerically. To compute integral $I_2$, we performed a Taylor expansion of function $G_3(q,h)$ in the vicinity of $h \simeq h_-(0)$.

Note again, similarly as for the velocity, that some terms still need to be dropped out in order to stay at third order in $q$ and that the calculation stands for $q \geq q_{\rm M}/2$, as long as we are happy with third order in $q$.
\label{lastpage}
\end{document}